\pdfoutput=1 

\documentclass[]{gpaalatex/gPAA2e}

\usepackage{makeidx}
\usepackage{multicol}
\usepackage{url}
\usepackage{algorithm}
\usepackage{algorithmicx}
\usepackage{algpseudocode}
\usepackage{listings}
\usepackage{cleveref}
\usepackage[parfill]{parskip}
\usepackage{wrapfig}
\usepackage{tikz}
\usepackage{standalone}
\usepackage[textsize=footnotesize]{todonotes}

\usetikzlibrary{calc}
\usetikzlibrary{shadows.blur}

\usepackage[utf8]{inputenc} 
\usepackage{xspace}         
\newcommand{\walberla}{\textsc{waLBerla}\xspace}
\newcommand{\TERRAN}{\textsl{HyTeG}\xspace}
\newcommand{\eg}{\mbox{e.\,g.}\xspace}
\newcommand{\ie}{\mbox{i.\,e.}\xspace}
\newcommand{\cf}{\mbox{cf.}\xspace}
\newcommand{\Fig}[1]{\mbox{Fig.\,\ref{#1}}}
\newcommand{\Sect}[1]{\mbox{Sec.\,\ref{#1}}}

\newcommand{\Alg}[1]{\mbox{Alg.\,\ref{#1}}}

\usepackage{pgfplots}
\definecolor{RYBred}{RGB}{225, 0,   0}
\definecolor{RYBdarkred}{RGB}{140, 0,0}
\definecolor{RYBblue}{RGB}{50, 120,   255}
\definecolor{RYBdarkblue}{RGB}{0, 0,140}
\definecolor{RYBSemiDarkRed}{RGB}{200, 0,0}
\definecolor{RYBDarkGreen}{RGB}{000, 130, 000}
\definecolor{RYBDarkCyan}{RGB}{000, 139, 139}
\definecolor{SlateGrey}{RGB}{112, 128, 144}
\definecolor{SlateBlue3}{RGB}{105, 089, 205}

\pgfplotscreateplotcyclelist{CyleList_Graphs}{%
   RYBdarkblue,line width=0.9pt,every mark/.append style={mark size = 2.8, fill opacity=0},mark=+\\%
   RYBDarkGreen,line width=0.9pt,every mark/.append style={mark size = 2.7, fill opacity=0},mark=x\\%
   RYBSemiDarkRed,line width=0.9pt,every mark/.append style={mark size = 2.7, fill opacity=0},mark=star\\%
   black,line width=0.9pt,every mark/.append style={mark size = 2.7},mark=asterisk\\%
   RYBDarkCyan,line width=0.9pt,every mark/.append style={mark size = 1.8, solid},mark=o\\%
   SlateBlue3, dashed,line width=0.9pt,every mark/.append style={mark size = 2.7, solid},mark=-\\%
   SlateGrey,densely dashed,line width=0.9pt,every mark/.append style={mark size = 2.0, solid, fill opacity=0},mark=|\\%
}

\algblockdefx{ParFor}{EndParFor}[1]%
{\textbf{for each }#1 \textbf{do in parallel}}%
{\textbf{end for}}

\begin{document}
\jvol{00} \jnum{00} \jyear{2018} 

\markboth{N. Kohl, D. Th\"onnes, D. Drzisga, D. Bartuschat, 
and U. R{\"u}de}{Parallel, Emergent and Distributed Systems}

\articletype{Article}

\title{ A Scalable and Modular Software Architecture for Finite Elements on Hierarchical Hybrid Grids }

\author{
Nils Kohl$^{\rm a}$, 
Dominik Th\"onnes$^{\rm a}$, 
Daniel Drzisga$^{\rm b}$,
Dominik Bartuschat$^{\rm a}$$^{\ast}$\thanks{$^\ast$Corresponding author. Email: dominik.bartuschat@cs.fau.de\vspace{6pt}},
and Ulrich R{\"u}de$^{\rm a,c}$
\\\vspace{6pt}  
$^{\rm a}${\em{Lehrstuhl f\"ur Systemsimulation, Friedrich-Alexander Universit\"at Erlangen-N\"urnberg, 91058 Erlangen, Germany}};\\
$^{\rm b}${\em{Institute for Numerical Mathematics (M2), Technische Universit{\"a}t M{\"u}nchen, 85748 Garching b. M\"unchen, Germany}};\\
$^{\rm c}${\em{Parallel Algorithms Group, CERFACS, 31057 Toulouse, France}}
\\\vspace{6pt}\received{received May 2018}
}

\maketitle

\begin{abstract}
In this article, a new generic higher-order finite-element framework for massively parallel simulations is presented.
The modular software architecture is carefully designed to exploit the resources of modern and future supercomputers.
Combining an unstructured topology with structured grid refinement facilitates high geometric adaptability and matrix-free multigrid implementations with excellent performance.
Different abstraction levels and fully distributed data structures additionally ensure high flexibility, extensibility, and scalability.
The software concepts support sophisticated load balancing and flexibly combining finite element spaces.
Example scenarios with coupled systems of PDEs show the applicability of the concepts to performing geophysical simulations.
\bigskip
\end{abstract}

\bigskip

\section{Introduction \label{sec:Intro}}
Current developments in computer architecture are driven by modern multi-processors
with an ever increasing parallelism.
This includes process-level parallelism between an increasing number of compute cores,
data-level parallelism in terms of vector processing (single instruction multiple data, SIMD),
and instruction-level parallelism.
Parallelism is even more essential in the area of scientific computing
where increasingly complex physical and technical problems
can be solved by numerical simulations on high performance computers.
With the advent of extreme-scale computing,
scientific software must be capable of exploiting the massive parallelism
provided by modern supercomputers.

In this article, the design and implementation
of a new finite element software framework \TERRAN (Hybrid Tetrahedral Grids) will be presented.
With \TERRAN, we aim to develop a flexible, extensible, and sustainable framework for massively parallel finite element computations.
To meet the requirements of a scalable and modular software structure,
a carefully designed software architecture is essential.
Our work builds on experience with designing complex and scalable software systems in scientific computing,
such as Hierarchical Hybrid Grids (HHG)~\cite{Bergen2003HHG, bergen2005LargestFEsystem,Gmeiner2016QuantitativeStokes}
and \walberla\footnote{\url{http://walberla.net}}~\cite{Feichtinger2011,kostler2013cse,Godenschwager2013,Schornbaum2015}.
Low-level hardware-aware implementation techniques are realised in a modular software architecture  
for extreme-scale efficiency on current and future supercomputing platforms.
In the scope of this article, the software concepts are presented for two-dimensional triangular finite elements.
All features, however, are implemented to be reusable also for three-dimensional simulations.

\subsection{Motivation}
\TERRAN is a generic framework for finite-element based computations building on some of the design principles of the preceding HHG software~\cite{Bergen2004HHGdatastruct},
but is designed to increase its functionality, flexibility, and sustainability.

\begin{figure}[h]
\centering
\subfigure[level 0\label{fig:hhg-refinement-level-0}]{
\resizebox*{0.28\textwidth}{!}{
\includegraphics[width=\textwidth]{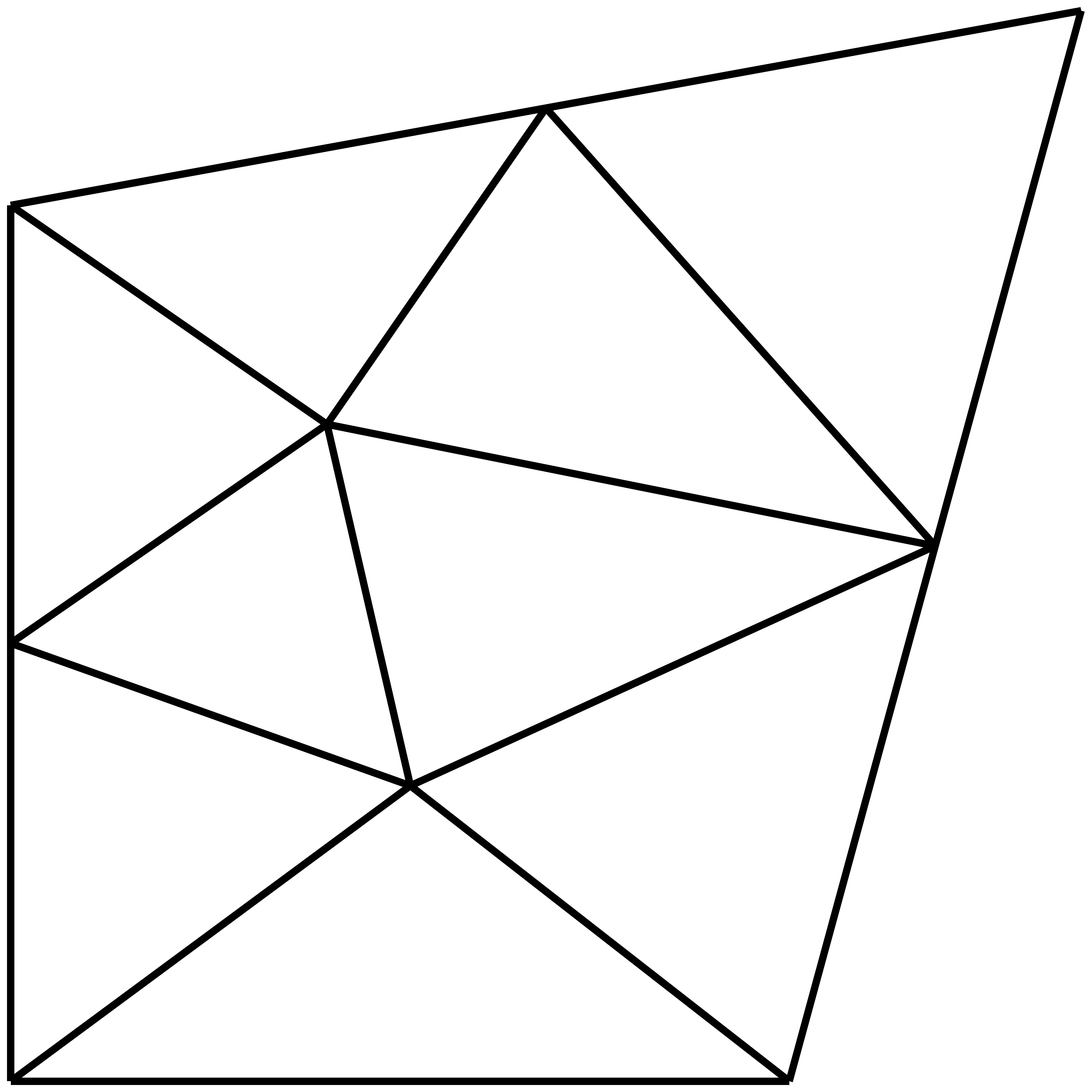}
}}%
\subfigure[level 1\label{fig:hhg-refinement-level-1}]{
\resizebox*{0.28\textwidth}{!}{
\includegraphics[width=\textwidth]{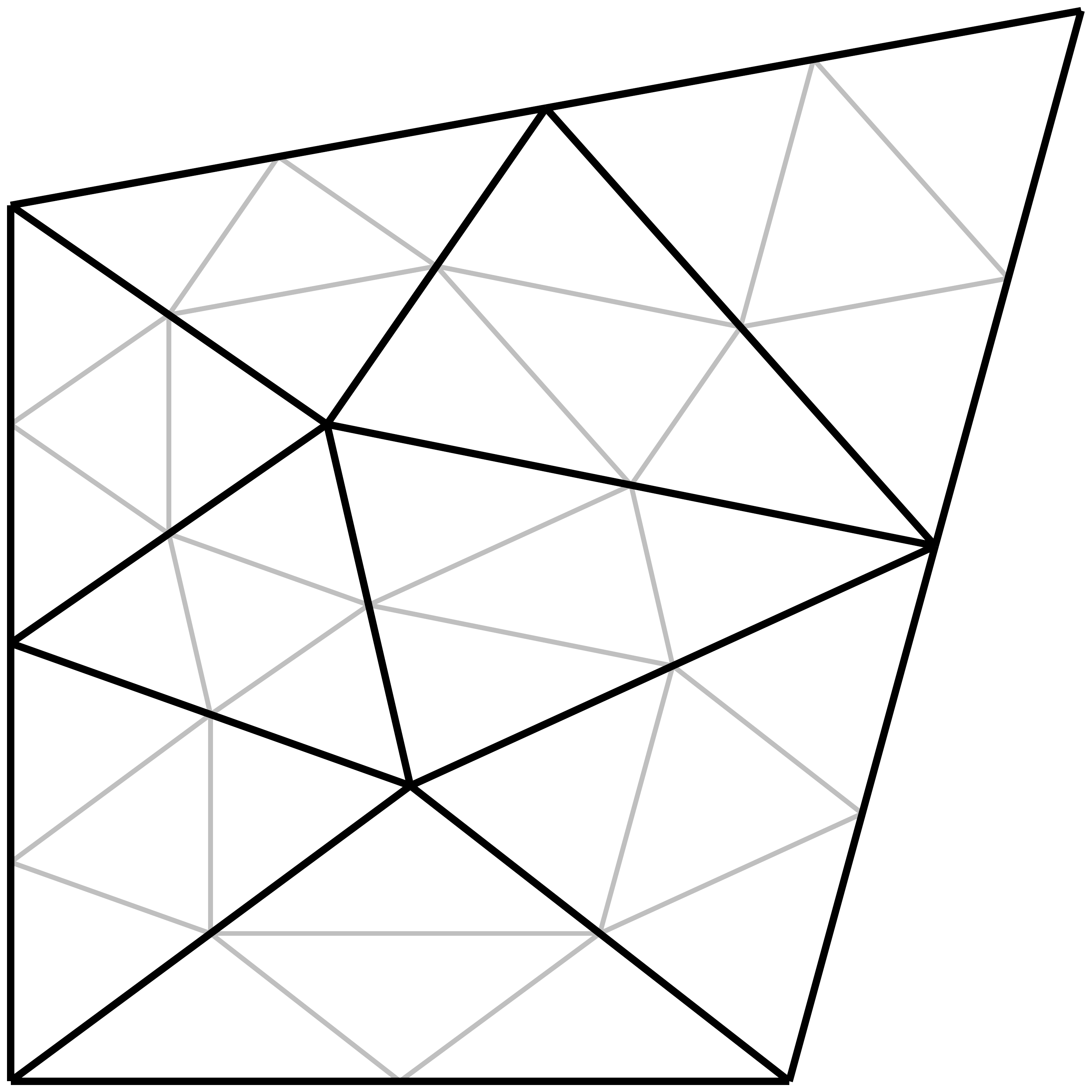}
}}%
\subfigure[level 2\label{fig:hhg-refinement-level-2}]{
\resizebox*{0.28\textwidth}{!}{
\includegraphics[width=\textwidth]{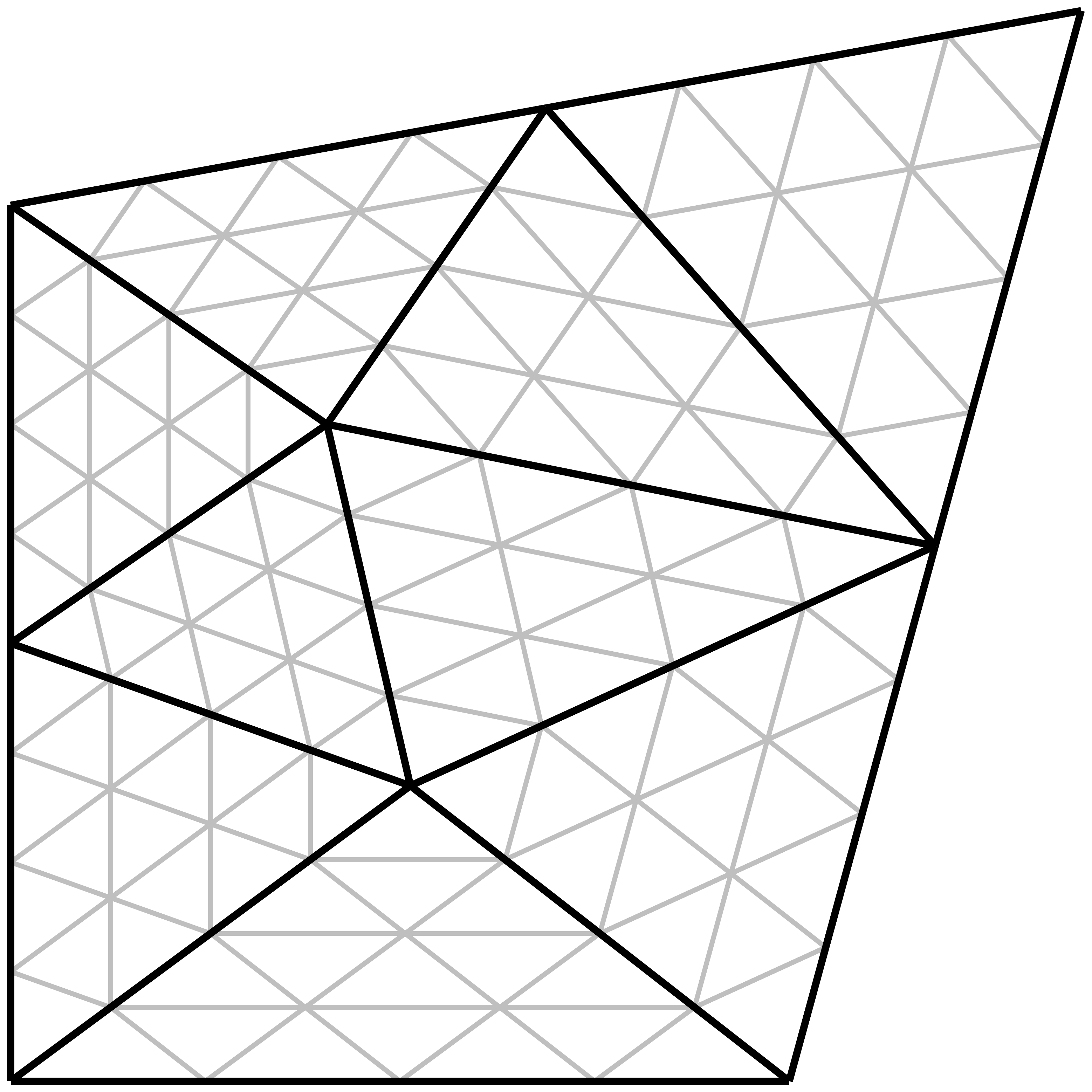}
}}%
\caption{Uniformly structured refinement of an unstructured mesh with triangular elements.
\Fig{fig:hhg-refinement-level-0} shows the unstructured mesh without refinement (referred to as refinement level 0). 
\Fig{fig:hhg-refinement-level-1} and \Fig{fig:hhg-refinement-level-2} depict structured refinement of level 1 and 2 applied to each triangular element.\label{fig:hhg-refinement}}
\end{figure}

Both frameworks build on a hierarchical discretisation approach that employs unstructured coarse meshes combined with structured and uniform refinement~\cite{Bergen2003HHG,Bergen2006Diss}. 
This structure is depicted in \Fig{fig:hhg-refinement} for a two-dimensional triangular mesh.
The resulting nested hierarchy of grids is used to implement geometric multigrid methods.
When progressing to extreme scale, multigrid methods are essential since they exhibit asymptotically optimal complexity. This algorithmic scalability makes them far superior to most alternative solver algorithms.
Additionally, the regularity of the refined elements is exploited by representing the system matrix in terms of \emph{stencils} whose entries correspond to the non-zero elements of a row in the stiffness matrix.
The matrix-free approach of HHG combined with a memory access structure that avoids indirections
is favourable for modern hardware, as it minimises both memory consumption
and memory access operations. 

Based on a similar approach, HHG was designed as a multigrid library with excellent efficiency for scalar elliptic problems \cite{Bergen2006MG_FE, bergen2005LargestFEsystem}
and for Stokes flow \cite{Gmeiner2015PerfomanceStokesHHG, Gmeiner2015PerfomanceStokesHHG2, Gmeiner2016QuantitativeStokes}.
The largest published results reach up to $10^{13}$ degrees of freedom (DoFs)~\cite{Gmeiner2016QuantitativeStokes},
exceeding the capability of alternative approaches by several orders of magnitude.
We emphasise that solving systems of such size
would not be possible on any current supercomputer 
with matrix-based methods where the sparse matrix must be stored explicitly.
Techniques like on-the-fly stencil assembly~\cite{Bauer:2016:OnTheFly,Bauer2017stencil} 
can keep the memory requirements low also for more complex problems.

HHG, however, is restricted to conforming linear finite elements only.
Therefore, \eg transport processes have to be modelled employing a duality
between nodal finite element meshes and polyhedral control volumes
for a finite volume discretisation, as in~\cite{waluga2016mass}.
The execution model is strictly bulk-synchronous and there is no
support for dynamic adaptivity and load balancing, significantly limiting the applicability. 
The new \TERRAN framework is therefore a completely new design built on similar principles
but generalizing the concept to remove these limitations.

\subsection{Design Goals and Contribution}
A core feature of the new \TERRAN code are multi-scale tetrahedral higher-order finite elements 
with a uniformly structured refinement.
This leads to excellent computational performance combined 
with high geometric flexibility and improved spatial accuracy for problems with sufficient regularity. 
The design as described here is restricted to simplices.
However, the software structure is kept flexible to also support hexahedral or prismatic elements in future versions.

The data layout of \TERRAN supports DoFs on the nodes, edges, faces, and volumes of a finite element mesh
to support most grid-based discretisations, including \eg discontinuous Galerkin methods.
Edge DoFs facilitate stable discretisations of flow problems as in 
the Taylor-Hood discretisation~\cite{elman2014finite}.

To achieve superior scalability, an improved \emph{domain partitioning} concept
with \emph{abstract data handling} is presented in \Sect{sec:topology}.
This software architecture
supports static and dynamic load balancing techniques, permits asynchronous execution, and it forms the
foundation for data migration,
advanced resilience techniques, and adaptive mesh refinement.
A newly introduced approach to classify and separately store the mesh data
based on their topological location in the finite element mesh is described in \Sect{sec:simdata}.
These mesh structures can be implemented with index-based memory access, avoiding 
the usual indirection of sparse matrix structures with their inherent performance penalty.
This direct access facilitates implementing the most time-consuming compute kernels with superior efficiency,
since these data structures are specifically designed
to better exploit instruction level parallelism and vectorisation.
An \emph{array-access abstraction} is presented that allows to adapt the 
underlying memory layout of the unknowns to the access patterns of the employed computational kernels.
Various solvers, but in particular geometric multigrid methods, can be implemented easily.
To this end, \TERRAN supports the possibility to combine 
different finite element spaces and the corresponding operators.
The algorithmic building blocks for coupled systems of partial differential equations (PDEs)
are described in \Sect{sec:numerics}.

Currently simulating Earth mantle convection is the primary application target
as a classical extreme-scale science problem.
Mantle convection is the driving force for plate tectonics, causing Earth quakes and mountain formation.
This application motivates the example computations presented in \Sect{sec:Results}.
However, the \TERRAN framework is also well suited for many other physical applications
that can make use of very large-scale finite element models.

\subsection{Related Work}
Many high performance simulation frameworks 
can be classified into two main categories based on the meshes 
they use for representing the simulation domain.
Frameworks that support finite elements on unstructured meshes include DUNE~\cite{Bastian2008DuneFramework,Bastian2008DuneImpl} with its module DUNE-FEM~\cite{Dedner2010DuneFem}, libMesh~\cite{Kirk2006libMesh},
UG4 \cite{Vogel2013ug}, and NEKTAR++~\cite{Cantwell2015NektarOverview}.
Another class of frameworks uses structured meshes, including 
deal.II~\cite{dealII85,BangerthHartmannKanschat2007}, 
Nek5000~\cite{Nek5000url}, and \walberla~\cite{Feichtinger2011}.

The frameworks DUNE, libMesh, UG4, and deal.II
support higher-order conforming and non-conforming finite elements as well as discontinuous Galerkin methods, combined with h-,p-,hp-refinement and adaptivity.
Additionally, DUNE and deal.II both provide element-based matrix-free methods on polyhedral elements and structured hexahedral grids, respectively.
NEKTAR++ is a tensor-product based high-order finite element package for tetrahedral, hexahedral, and prismatic elements that
employs tensor-product approximations to significantly reduce computational costs~\cite{Cantwell2015NektarOverview}.
Nek5000 is a highly scalable spectral element code based on hexahedral elements that currently supports h-adaptive mesh refinement \cite{Peplinski2016hRefNek5000}.

Unstructured meshes have the advantage of geometric flexibility,
however, the implementation is often less efficient and can at this time not reach simulation
sizes as \eg~demonstrated in \cite{Gmeiner2016QuantitativeStokes,Gmeiner2015PerfomanceStokesHHG,Gmeiner2015PerfomanceStokesHHG2}.
Structured meshes support matrix-free methods that 
can be used to save memory and memory access bandwidth.
They are also better suited to exploit the microarchitecture of modern processors,
in particular accelerators, such as GPUs.
The \TERRAN framework presents a compromise between
structured and unstructured meshes trying to preserve the geometric 
flexibility of unstructured meshes while still
providing the efficiency that can be achieved by hardware-aware implementations 
on structured meshes.

\section{Domain Partitioning and Subdomain Mesh Topology\label{sec:topology}}
An efficient and fully distributed data structure is a vital component for scalable parallel
simulation software. In this section, a scheme and the corresponding data structures in \TERRAN are presented to partition an unstructured 
mesh as it is used for hybrid discretisations as introduced in \Sect{sec:Intro}.

\subsection{Domain Partitioning\label{subsec:domainpartitioning}}
Parallel implementations of stencil-based update rules require
a domain partitioning strategy that allows for read-access to the neighbourhood of an unknown.
The data dependencies across process boundaries are typically resolved by extending the local subdomains by \emph{ghost layers}, also referred to as \emph{halos}, 
that represent read-only copies of unknowns beyond the boundary of the subdomains.
Data points on the ghost layers are communicated across the subdomain boundaries
between two subsequent iterations.
The hierarchical partitioning scheme of the unstructured coarse grid presented in this section is motivated by the special properties 
of finite element discretisations on meshes that result from the structured refinement of unstructured meshes \cite{Bergen2004HHGdatastruct}.
This scheme separates subdomains with different local stencils and results in a unique assignment of unknowns to processes.

\TERRAN introduces the concept of {\em macro-primitives}.
A macro-primitive describes a geometrical object, its position, and orientation.
There are different types of macro-primitives that correspond to their respective topological relations: macro-{\em faces} for two-dimensional objects (\eg~triangles),
macro-{\em edges} for one-dimensional lines, and macro-{\em vertices} for points.

\begin{figure}[h]
\centering
\resizebox{.85\linewidth}{!}{\includegraphics{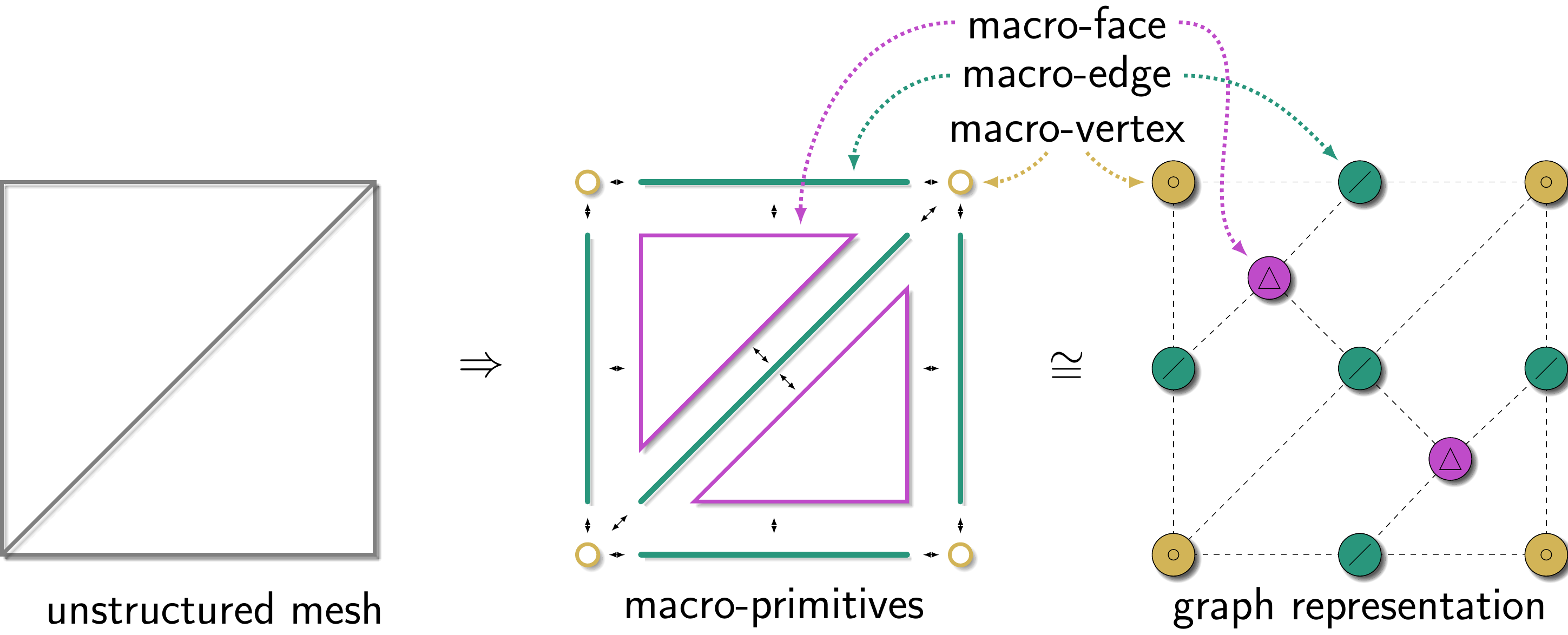}}
\caption{Schematic illustration of the internal macro-primitive graph representation of a 2D example mesh.}
\label{fig:topology}
\end{figure}

The unstructured mesh is conceptually converted into a graph of these
macro-primitives as illustrated in \Fig{fig:topology}.
The graph's vertices are instances of macro-primitives.
To construct the graph, one graph-vertex that represents one macro-primitive per mesh-vertex, mesh-edge, and mesh-face of the unstructured mesh is inserted into the graph.

The graph-edges connect the graph-vertices to reflect the topology of the unstructured mesh.
Macro-primitives are connected with their neighbours of the next lower and higher dimension, \ie, 
all graph-vertices of type macro-face are only connected to the adjacent macro-edges but not to neighbouring macro-faces.

\subsection{Simulation Data\label{sec:datacontainer}}
The macro-primitives also serve as containers for arbitrary data structures.
With this abstract data-handling, any kind of data can be attached to a macro-primitive, including instances of custom classes or standard C++ data structures.

\begin{figure}[ht]
\centering
\resizebox{.85\linewidth}{!}{\includegraphics{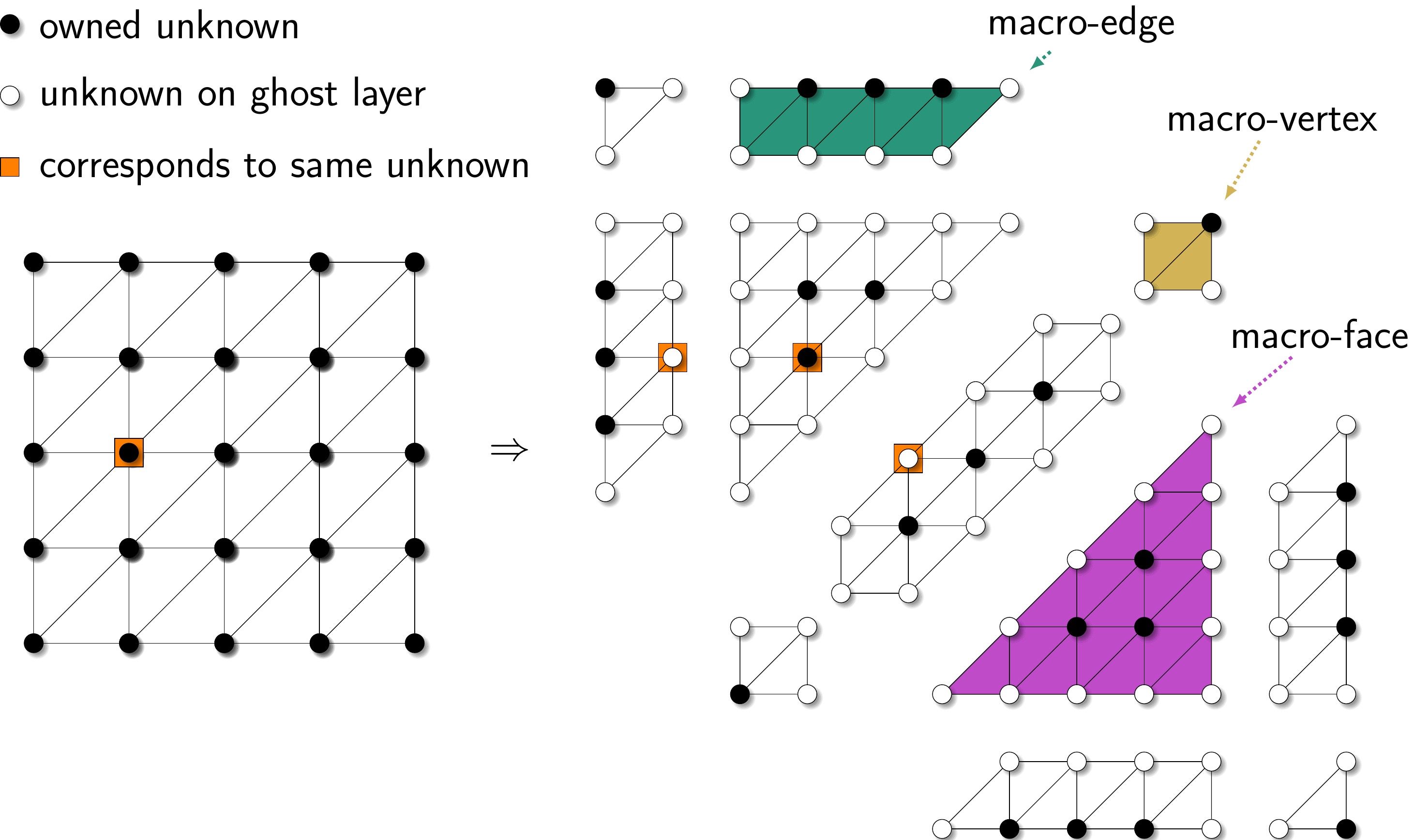}}
\caption{Schematic assignment of the unknowns of a P1 finite-element discretisation to the macro-primitives after the domain partitioning process.}
\label{fig:halos}
\end{figure}

To construct the hierarchy of refined subdomain meshes,
data structures that represent fields (\ie arrays) 
of unknowns are allocated and attached to the macro-primitives.
The size and shape of the fields on each macro-primitive depend on the primitive's geometry, the neighbouring macro-primitives, and the refinement level.
The fields are extended by ghost layers to facilitate data exchange
with the neighbouring macro-primitives when these are stored on different processors in a distributed memory system.
\Fig{fig:halos} illustrates the structure of such fields of unknowns that arise from a P1 finite-element discretisation on a mesh that is refined by two levels, and is generated based
on two coarse-grid mesh elements,  
corresponding to the mesh in \Fig{fig:topology}.
Each unknown is assigned to exactly one macro-primitive. The ownership is illustrated by black points in \Fig{fig:halos}.
A read-only copy may reside in the ghost layers of other macro-primitives. The ghost layers are illustrated by white points.
The orange squares indicate the same unknown that is only owned by one macro-primitive but resides as a read-only copy in the ghost layers of neighbouring macro-primitives.

The container approach facilitates a flexible extension of the framework by allowing to attach arbitrary fields to the macro-primitives--especially fields 
that represent unknowns of different discretisations, \eg higher-order finite element or finite volume discretisations. 
The field data-structure is discussed in more detail in \Sect{sec:simdata}.

Moreover, the macro-primitives can be attributed with
arbitrary data and are not restricted to field data-structures.
Such data structures could also for example store logistic data as required for handling
special boundary conditions or metadata that is needed for load balancing.
Similar approaches to decouple the simulation data from the coarse-grained topology have been presented in \cite{Schornbaum2015,Acun2014} 
and have been shown to provide an elegant interface for runtime load balancing \cite{Schornbaum2017}, resilience, and data migration \cite{Kohl2017}.
The macro-primitive graph data structure is therefore not restricted to finite-element based simulation techniques.

To realise runtime load balancing or checkpointing techniques, data serialisation and migration to other processes or to permanent storage are required.
Therefore, an interface is provided for callback functions that (de-)serialise the attached data structures.
The framework is then able to (de-)serialise and store or migrate the attached data (\eg fields of unknowns) without knowledge of its actual internal structure, which allows a decoupled design of the migration process.

\subsection{Load Balancing}
The graph of macro-primitives is the basic data structure to distribute the domain among different processes.
During the partitioning of the unstructured mesh to a graph of macro-primitives, each macro-primitive is attributed with a globally unique identifier and is assigned to one process.
Each process, however, may own multiple macro-primitives.
The connectivity is stored in a distributed fashion as each macro-primitive stores the identifiers of its neighbouring macro-primitives according to the graph's structure.
As a result, no global information about the domain is stored on any process and the mesh is completely distributed.
Thus, arbitrary sized meshes can be processed given a sufficient number of processes---this is essential for the design of simulation software that can reach extreme scale.

Simulation on a large partitioned domain often
requires a flexible and dynamic load balancing concept.
Approaches to create and balance partitions often rely on a graph representation of the targeted domain.
The macro-primitive graph shown in \Fig{fig:topology} represents such a structure.
Each node of the graph represents one macro-primitive of the domain while the graph-edges reflect the communication paths.

\begin{figure}[h]
\centering
\subfigure[macro-faces \label{fig:loadbalancing:macrofaces}]{
\resizebox*{0.85\textwidth}{!}{
\includegraphics[width=\textwidth]{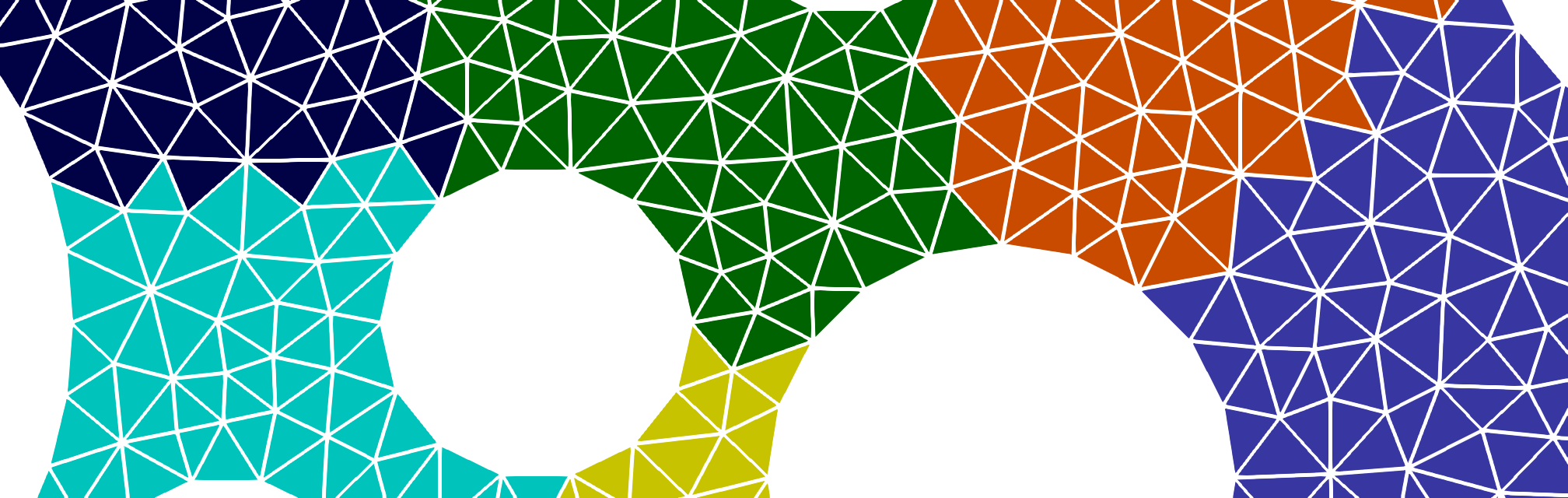}
}}\\%
\subfigure[macro-edges \label{fig:loadbalancing:macroedges}]{
\resizebox*{0.85\textwidth}{!}{
\includegraphics[width=\textwidth]{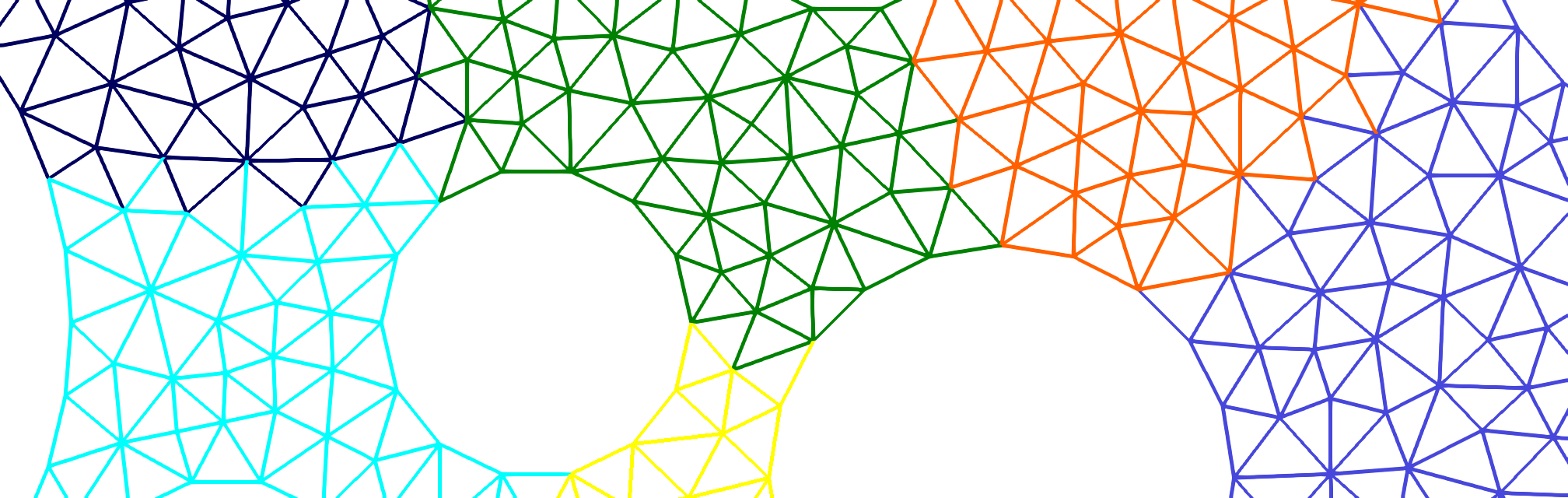}
}}\\%
\subfigure[macro-faces with structured refinement \label{fig:loadbalancing:refined}]{
\resizebox*{0.85\textwidth}{!}{
\includegraphics[width=\textwidth]{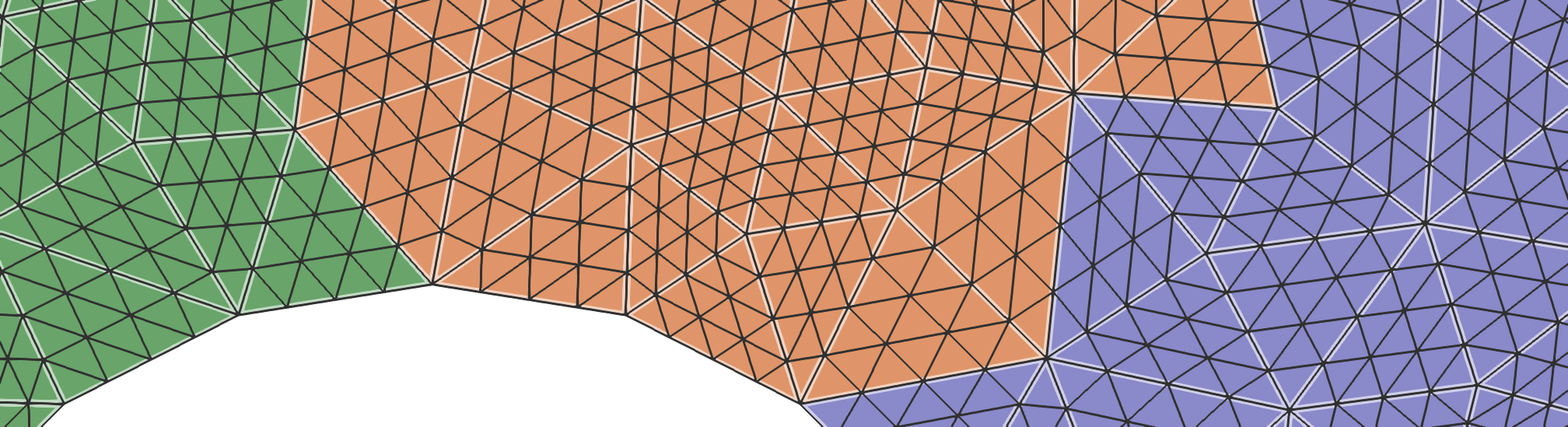}
}}%
\caption{Section of a balanced macro-primitive mesh, with colouring indicating the target process. The individual figures emphasise that not only the macro-faces (\cf~\Fig{fig:loadbalancing:macrofaces}) 
but also the macro-edges (\cf \Fig{fig:loadbalancing:macroedges}) and macro-vertices are individually assigned to processes.
\Fig{fig:loadbalancing:refined} illustrates the structured refined mesh on refinement level~2.}
\label{fig:loadbalancing}
\end{figure}

Therefore, general graph partitioning algorithms can be applied to this structure without knowledge of the underlying application.
Well-designed load balancing libraries already exist \cite{Karypis1998,PTScotch2008,Zoltan2012} and can be used to create partitions of the macro-primitive graph.
\Cref{fig:loadbalancing} shows a two-dimensional mesh containing some circular obstacles that was balanced among eight processes using ParMetis \cite{Karypis1998}.

Such partitions can be created statically during a setup phase or dynamically at runtime.
Dynamic load balancing requires runtime data migration and can be realised in a two step process. First, the current partitioning of the global graph is adapted. This could by done by an external library such as ParMetis~\cite{Karypis1998}.
Then the simulation data is migrated to the respective processes. In \TERRAN, the migration process is realised using the (de-)serialisation routines as 
described in \Sect{sec:datacontainer}.
A scalable dynamic load balancing concept applied to simulations on adaptive meshes is presented in~\cite{Schornbaum2017}.

The partitioning algorithms can be augmented by weights that are assigned to the individual graph-nodes and -edges.
In a finite-element context, a node weight could represent the number of unknowns located in the respective primitive and therefore serves as indicator of the corresponding work load.

Due to the connectivity of the macro-primitive graph, macro-edges are likely to end up on the same processor as their neighbouring macro-faces and vice versa.
Since most communication during the simulations takes place along the edges of the macro-primitive graph, this structure naturally yields suitable distributions if edge-cut minimizing partitioning algorithms are employed.

\subsection{Communication\label{sec:topocomm}}

Synchronisation of the read-only copies of unknowns on the ghost layers is employed through communication between neighbouring macro-primitives.
The macro-primitive graph structure illustrated in \Fig{fig:topology} reflects the direct neighbourhood of the primitives, \ie, communication is only performed along the graph-edges.
For example, the ghost layers of macro-faces are updated by communication with the neighbouring macro-edges.

The communication pattern is highly dependent on the update pattern performed on the unknowns and must be adjusted individually.
Explicit update rules like matrix-vector products or Jacobi-type smoothing iterations can be combined with more efficient communication patterns than implicit update rules like for example Gauss-Seidel-type smoothing iterations.
In general, the latter require a certain global update succession.
More details are discussed in \Sect{sec:numerics}.

Corresponding to the \emph{abstract data-handling }concept to attach arbitrary data structures to the macro-primitives, the framework supports a similarly flexible communication abstraction.
A three-layer abstraction is employed to decouple the individual components of the communication process from another.

The \emph{buffer layer} provides abstraction from the actual (MPI-)send and receive calls and the internal data buffer structure.
C++ operator overloading for STL data structures and basic data types allows for a convenient (de-)serialisation of the transferred data.

The \emph{packing layer} is responsible for the (de-)serialisation of the attached data of a macro-primitive from and to data buffers.
It provides a (de-)serialisation interface that is implemented for each data structure that must be communicated.
For example, there are implementations for (de-)serialisation of unknowns that shall be sent to the ghost layers of neighbouring primitives.
Upon invocation, the serialisation routines pack the respective unknowns into a buffer on one primitive and the deserialisation routines unpack them on the ghost layers of the corresponding neighbour.
However, the packing layer does not employ any communication but is only called to (un-)pack data from and to buffers.

The \emph{control layer} manages the communication directions along the graph-edges of the macro-primitive graph.
Before sending and after receiving buffers, it calls the respective (de-)serialisation routines of the packing layer without knowledge of the actual data structures.
The control layer employs optimisations like non-blocking communication and calls process-local communication routines if graph-edges do not cross process boundaries.

This decoupled design allows for intensive code reuse and enhances the extensibility.
The control layer is not affected by the introduction of new data structures since only the interface of the packing layer must be implemented to allow for communication of the respective data.
A well-tested buffer layer abstraction is for example already implemented in the core of the \walberla framework \cite{Feichtinger2011} which serves also 
as a basis for this implementation.

\section{Structured Refinement of Mesh Data
\label{sec:simdata}}
In this section, the structured field data-structures for
the DoFs inside the macro-primitives are presented
that result from the structured refinement of an unstructured topology.
Since the largest part of the unknowns is located on the macro-faces the descriptions in the following are focused on these primitives. 
The presented concepts can be adapted for macro-edges and macro-vertices as well.

\subsection{Structured Subdomains}
As illustrated in \Fig{fig:hhg-refinement}, starting from a triangulation of the domain, a recursive refinement is performed by connecting the
three midpoints of the triangle edges. This results in four new triangles of the same congruence class as \eg presented
in \cite{bank1980refinement}. 

\begin{figure}[ht]
\centering
\subfigure[vertex DoFs \label{fig:blank_vertex}]{
\resizebox{0.29\textwidth}{!}{
\includegraphics[width=\textwidth]{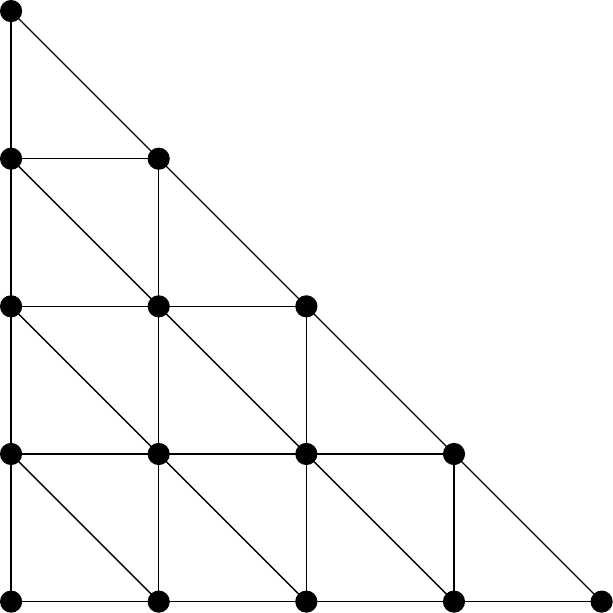}
}}%
\subfigure[edge DoFs \label{fig:blank_edge}]{
\resizebox{0.29\textwidth}{!}{
\includegraphics[width=\textwidth]{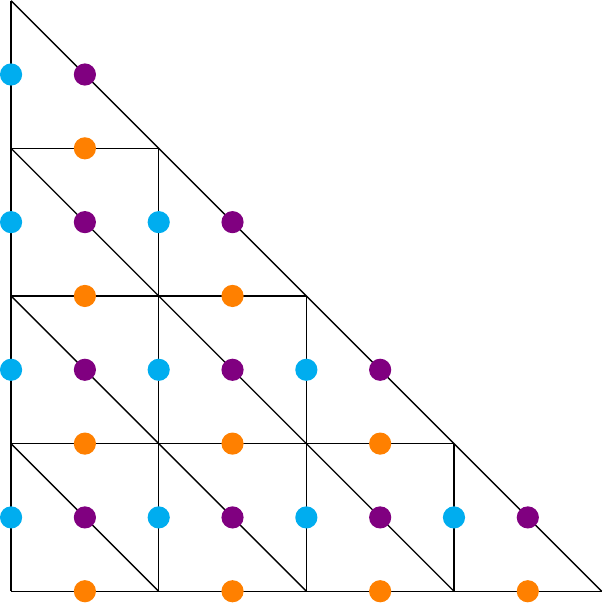}
}}%
\subfigure[face DoFs \label{fig:blank_cell}]{
\resizebox{0.29\textwidth}{!}{
\includegraphics[width=\textwidth]{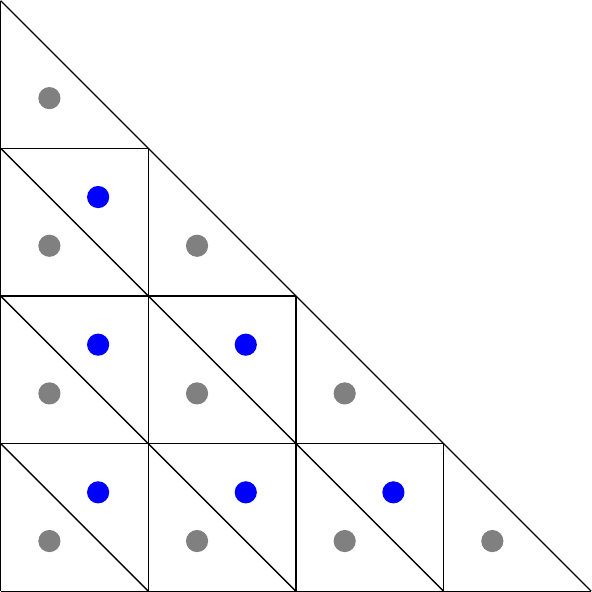}
}}%
\caption{Different types of degrees of freedom placed on various positions on the mesh. The colours denote the subgroups
  of the individual types depending on the topological location of the DoFs on the elements.}
\label{fig:different_dofs}
\end{figure}

Unknowns are placed at certain positions in the structured refined element mesh depending on the corresponding finite element or finite volume discretisation. 
They are classified as \emph{vertex-unknowns}, \emph{edge-unknowns} and \emph{face-unknowns} reflecting their topological position on the mesh, \cf \Fig{fig:different_dofs}.

In the case of first-order finite elements, the unknowns are placed on the vertices as shown in \Fig{fig:blank_vertex}.
Second-order finite elements additionally require DoFs on the edges as displayed in \Fig{fig:blank_edge}.
Other discretisations, such as \eg finite volumes can be realised by placing the unknowns on the interior of the triangles as illustrated in \Fig{fig:blank_cell}. 
More details are given in \Sect{subsec:discretizations}.

The edge-unknowns are further separated into subgroups depending on whether they are placed on a horizontal, vertical or
diagonal edge of an element. The different groups are shown in different colours in \Fig{fig:blank_edge}.  A similar grouping for
rectangular elements has been presented in \cite{kuckuk2016automatic}. 

Similar to the edge-unknowns, the face-unknowns are also separated into subgroups depending on the orientation of the
element they reside in. As illustrated in \Fig{fig:blank_cell}, there are two subgroups: \emph{upward-} \emph{upward-} and \emph{downward-facing} elements.

One subgroup on its own (\eg all horizontal edge unknowns or all face-unknowns in upward-facing elements) leads again to a triangular pattern similar to that of the vertex unknowns. 
Because of the similar patterns, a single array-access calculation routine for variable sized triangular field layouts is capable of calculating the array accesses for vertex-unknowns as well as for each subgroup of edge- and face-unknowns. Therefore, the separation into subgroups avoids code duplication in the sense that the array-access routines can be reused for all types of unknowns.

Instead of modularizing  access patterns and memory layouts by element type, the required features are implemented per type of
unknown, \ie, modules are implemented for vertex-unknowns, edge-unknowns, and face-unknowns.  By combining these modules,
arbitrary element types can be created, for example finite elements of different order. This approach reduces
complexity and code duplication since routines that process vertex-unknowns can be reused for all
discretisations that require vertex-unknowns. 
Possible examples include numerical routines that update the unknowns or routines that serialise data for communication or IO. 
Additionally, advanced applications that employ advanced or 
uncommon element types can be prototyped 
and implemented rapidly by combining different types of unknowns.

\subsection{Indexing
  \label{subsec:indexing}}

\begin{figure}
\centering
\subfigure[topological indexing \label{fig:vertex_coords}]{
\resizebox*{0.44\textwidth}{!}{
\includegraphics{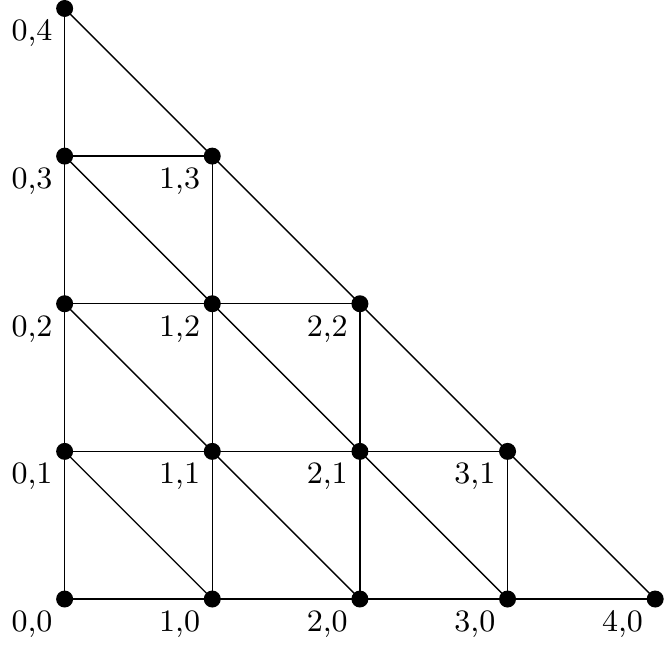}
}}%
\subfigure[underlying memory layout \label{fig:vertex_index}]{
\resizebox*{0.44\textwidth}{!}{
\includegraphics{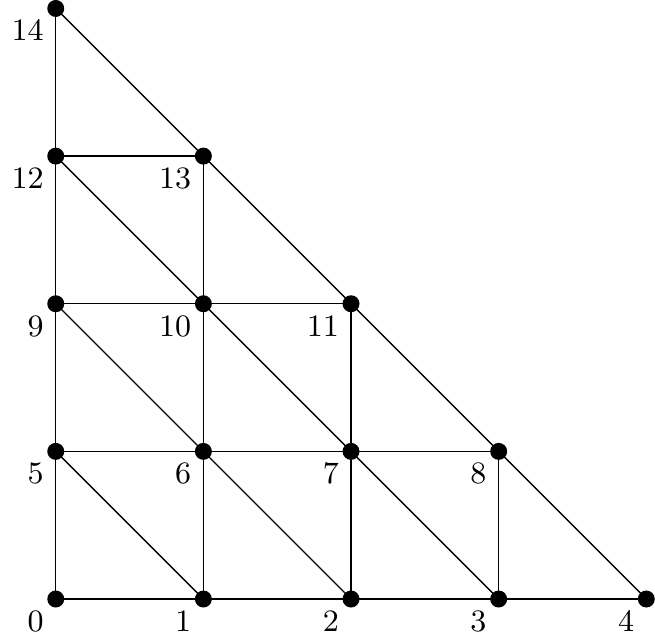}
}}%
\caption{Mapping from topological indexing of unknowns located at micro-vertices to one possible memory layout \label{fig:logicalcoordinates}}
\end{figure}

One challenge when comparing these triangular fields with their equivalents on quadratic or rectangular meshes (as \eg
used in \cite{dealII85}) is the translation of the topological index to the actual memory layout. While in the rectangular
case the number of unknowns is constant in each row and column, this is not the case for triangles, \cf \Fig{fig:logicalcoordinates}. 
In case of a simple linear data arrangement as shown in \Fig{fig:vertex_index}, this means that the offset from one row to another
decreases constantly.

Furthermore, it is desirable that the memory layout can be changed
to fit the requirements of a compute kernel and to
improve the performance for the specific processor microarchitecture.
Possible scenarios could be colouring schemes \eg used in a multi-colour Gauss-Seidel smoother or the possibility to use SIMD operations more efficiently.
For both of these reasons a layer of abstraction is introduced to separate the indexing from the actual memory layout.
A fixed topological enumeration is introduced and flexible methods are implemented to access the unknowns using indices.

An example for DoFs located on the vertices on a macro-face can be seen in \Fig{fig:logicalcoordinates}. 
The fixed topological indexing is shown in \Fig{fig:vertex_coords}. 
These coordinates are translated via an \emph{indexing function} into one possible memory layout as shown in \Fig{fig:vertex_index}.
The memory layout can be exchanged according to the specific needs of the application simply by exchanging the indexing function. 
The topological enumeration is defined for each type of unknown, \ie vertex-, edge- or face-unknowns.

By introducing the abstract indexing, the array access calculations are decoupled from the algorithms that access the
unknowns. Since the translation to absolute array indices is hidden behind the indexing functions, array layouts can be
switched by simply exchanging the indexing function. Consequently, different memory layouts can be added during later
phases of the development and can be compared to each other without rewriting existing kernels that are not performance
critical while time consuming compute kernels can be tuned for the specific layout. 
This provides the possibility to employ efficient
memory access patterns to improve \eg cache locality by using techniques as described in
\cite{kowarschik2000cache} and \cite{kowarschik2002performance}.

\subsection{Inter-Primitive Data Exchange}
To update the ghost layers of the fields of unknowns as introduced in \Sect{subsec:domainpartitioning} and illustrated in \Fig{fig:halos},
communication between neighbouring primitives must be employed. 

Due to the underlying unstructured mesh, it is not possible to implicitly calculate the neighbourhood relations of a certain macro-primitive.
The number of neighbours may vary for each primitive type. While macro-faces always
have three neighbouring macro-edges and macro-vertices, a macro-edge can have one or two neighbouring macro-faces. A
macro-vertex can have an arbitrary number of neighbouring primitives depending on the mesh. Since the data structures are
fully distributed there is no global knowledge about the neighbourhood relations. Therefore, local neighbourhood information
must be explicitly stored for each primitive.

As explained in \Sect{sec:topocomm} a layered approach is used where the control layer takes care that the correct
primitives communicate with each other. Before sending the data, however, serialisation is needed as well as packing the
data into a buffer that is sent to a receiving primitive belonging to a process located on a remote node. This means
that each macro-face, macro-edge, and macro-vertex needs to copy the data that all of their neighbours require into one or more buffers. 
If two adjacent macro-primitives are located on the same process, the data can be copied directly from one array to the other without (de-)serialisation to or from a buffer.

It is important to point out that in case of a macro-face the memory access pattern is quite different for each of the
three neighbouring macro-edges. The desired data might be located consecutively in memory but there could also be varying strides between
the entries.
Once more, the indexing abstraction discussed in
\Sect{subsec:indexing} can be used to simplify the
corresponding serialisation kernels.

Another complication is the ordering of unknowns in the topological layout. Since the input mesh is fully unstructured it is
not guaranteed that the orientation of one side of a macro-face is equivalent to the orientation of the neighbouring
macro-edge. This means that the particular order of unknowns might have to be adjusted during the communication. The
approach to solving this problem is that the primitive of higher dimension takes care of possible adjustments. For example,
a macro-face would write the data in reverse order into the send buffer such that the corresponding macro-edge can simply read the
data out of the buffer consecutively.

The steps that have to be performed when communicating one side of a macro-face to the adjacent macro-edge are as follows:
\begin{enumerate}
\item Determine which of the three sides of the macro-face the macro-edge is adjacent to and how the topological layout is oriented.
\item Determine the process associated with the adjacent macro-edge.
  \begin{enumerate}
  \item[(a)] If the edge is located on the same process: copy the macro-face data directly into the ghost layers of the macro-edge considering the orientation.
  \item[(b)] If the macro-edge is located on a different process:
    \begin{enumerate}
    \item[(i)] copy the macro-face data into the send buffer considering the orientation,
    \item[(ii)] transfer the buffer to the process the macro-edge is located on, and
    \item[(iii)] copy the data from the receive buffer into the ghost layers of the macro-edge.
    \end{enumerate}
  \end{enumerate}
\end{enumerate}

\begin{figure}[h]
\centering
\resizebox{.92\linewidth}{!}{\includegraphics{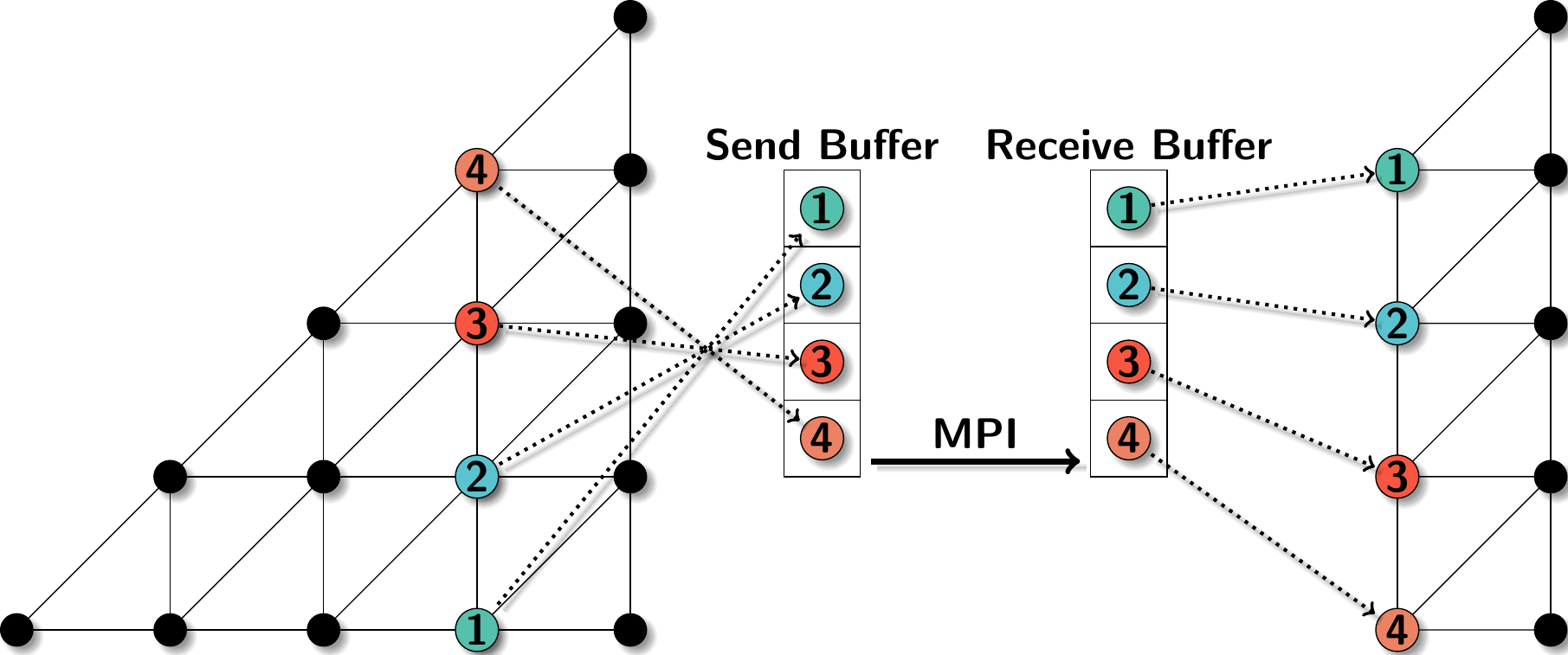}}
\caption{Communication of one side of the macro-face to the corresponding macro-edge. The example is chosen such that the order of the unknowns must be adjusted on the sender side during the data exchange}
\label{fig:data_comm}
\end{figure}

In \Fig{fig:data_comm} an example communication from one macro-face to one of the three neighbouring macro-edges is depicted. 
In this case, the unknowns are written to the buffer in reversed order by the macro-face since the orientation in the macro-edge
is opposite.

\section{Numerical Methods and Linear Algebra\label{sec:numerics}}
The discretisation of PDEs on the mesh structures of \TERRAN typically leads to large systems of equations.
For the implementation of iterative solvers, standard operations like addition, scalar products, and matrix-vector multiplications have to be performed. 
In the following, the realisation of these concepts within \TERRAN is described.

\subsection{Discretisations\label{subsec:discretizations}}
\begin{figure}
   \centering
\subfigure[P1 element \label{fig:elementP1}]{
\resizebox*{0.25\textwidth}{!}{
\includegraphics{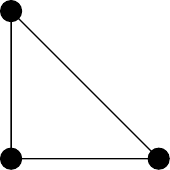}
}}%
\subfigure[P2 element \label{fig:elementP2}]{
\resizebox*{0.25\textwidth}{!}{
\includegraphics{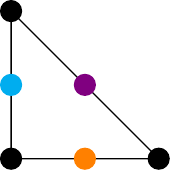}
}}%
\subfigure[P3 element \label{fig:elementP3}]{
\resizebox*{0.25\textwidth}{!}{
\includegraphics{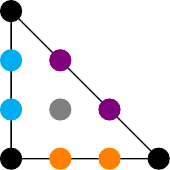}
}}%

   \vspace{12pt}
\subfigure[FV/DG0 element \label{fig:elementDG0}]{
\resizebox*{0.25\textwidth}{!}{
\includegraphics{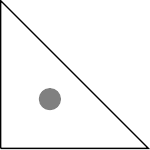}
}}%
\subfigure[DG1 (C) element \label{fig:elementDG1Cell}]{
\resizebox*{0.25\textwidth}{!}{
\includegraphics{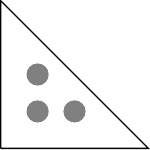}
}}%
\subfigure[DG1 (E) element \label{fig:elementDG1Edge}]{
\resizebox*{0.25\textwidth}{!}{
\includegraphics{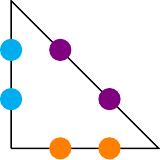}
}}%
   \caption{Conforming finite elements P1 (a), P2 (b), and P3 (c). Non-conforming finite volume (FV) or discontinuous Galerkin (DG) element of zeroth order (d), cell-based DG of first order (e) or edge-based DG of first order (f).}
   \label{fig:finiteelements}
\end{figure}

The data structures described in \Sect{sec:simdata} support storing unknowns at different positions in the topology of the underlying mesh.
By grouping different unknowns together, finite elements or other discretisations can be constructed.
\Fig{fig:finiteelements} illustrates some examples of typical finite elements. 
The first row from \Fig{fig:elementP1}~to~\Fig{fig:elementP3} presents the usual
conforming finite elements \cite{elman2014finite} from first to third order. 
Additionally, also non-conforming finite elements that are discontinuous across the edges can be realised, \cf~\Fig{fig:elementDG0}~to~\Fig{fig:elementDG1Edge}. 
The elements in Figs.~\ref{fig:elementDG1Cell} and \ref{fig:elementDG1Edge} are inherently equivalent but differ only in the location on the element the unknowns are associated with.

A global variable on the mesh discretised by a specific element may be represented by the union of all unknowns. 
This union represents a \emph{vector} in a finite-dimensional linear space.
On these vectors, the \TERRAN framework implements a set of functions similar to the BLAS level~1 routines~\cite{Lawson1979}. 
For example, the routine \texttt{assign} shown in \Alg{algo:assign}
computes a linear combination of two vectors.

\begin{algorithm}[h]
\caption{Assign $x_{dst} := \alpha x_{src_1} + \beta x_{src_2}$ without communication}
\label{algo:assign}
\begin{algorithmic}[1]
	\Procedure{assign}{$\alpha, x_{src_1}, \beta ,x_{src_2}, x_{dst}$}
	\ParFor{primitive}
	\State on primitive compute locally $x_{dst} := \alpha x_{src_1} + \beta x_{src_2}$
	\EndParFor
	\EndProcedure
\end{algorithmic}
\end{algorithm}

Similarly, it is possible to implement reductions on such vectors. As an example, the implementation of the equivalent of the \texttt{xDOT} BLAS routine is presented in \Alg{algo:dot}.

\begin{algorithm}[h]
   \caption{Dot product $dot := x_{src_1}^\top\,x_{src_2}$ with global reduction}
   \label{algo:dot}
   \begin{algorithmic}[1]
      \Procedure{dot}{$x_{src_1}, x_{src_2}$}
      \ParFor{primitive}
      \State on primitive compute $dot_{\text{local}} := x_{src_1}^\top\,x_{src_2}$
      \EndParFor
      \State compute by a global reduction the sum of all $dot_{\text{local}}$ and save it in $dot$
      \State \Return $dot$
      \EndProcedure
   \end{algorithmic}
\end{algorithm}

Note that the computation performed within a primitive is independent of others,
thus each primitive may be processed in parallel.

\subsection{Matrix-Free Approach\label{subsec:matrixfree}}
\begin{figure}[h]
\centering
\subfigure[Vertex DoF\label{fig:vertexDof}]{
\scalebox{0.55}{%
\includegraphics{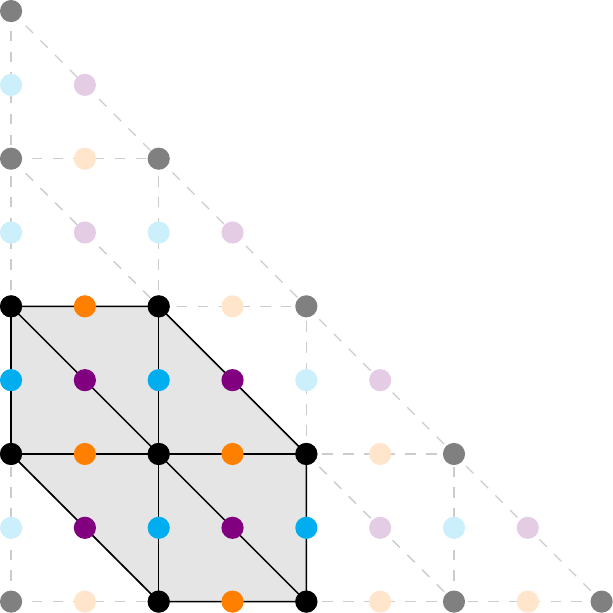}
}}\\[6pt]
\subfigure[Diagonal edge DoF\label{fig:diagEdgeDoF}]{
\scalebox{0.55}{%
\includegraphics{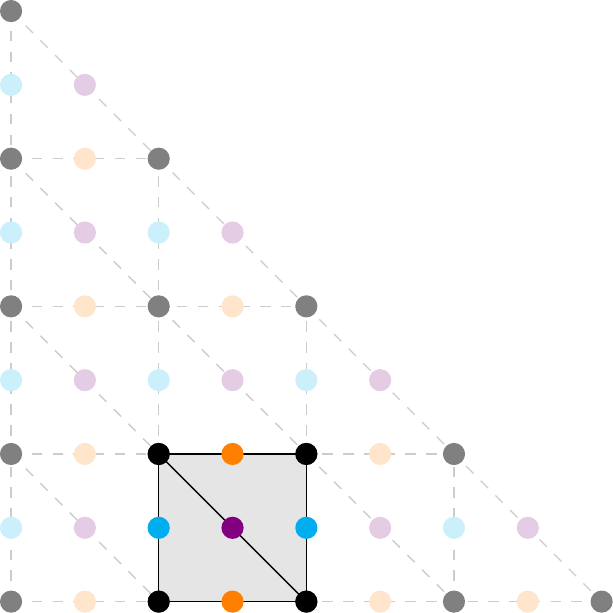}
}}
\subfigure[Horizontal edge DoF\label{fig:horEdgeDoF}]{
\scalebox{0.55}{%
\includegraphics{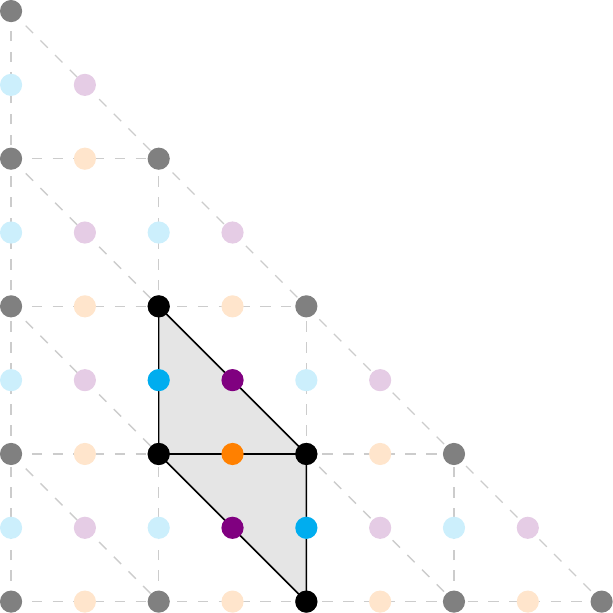}
}}
\subfigure[Vertical edge DoF\label{fig:vertEdgeDoF}]{
\scalebox{0.55}{%
\includegraphics{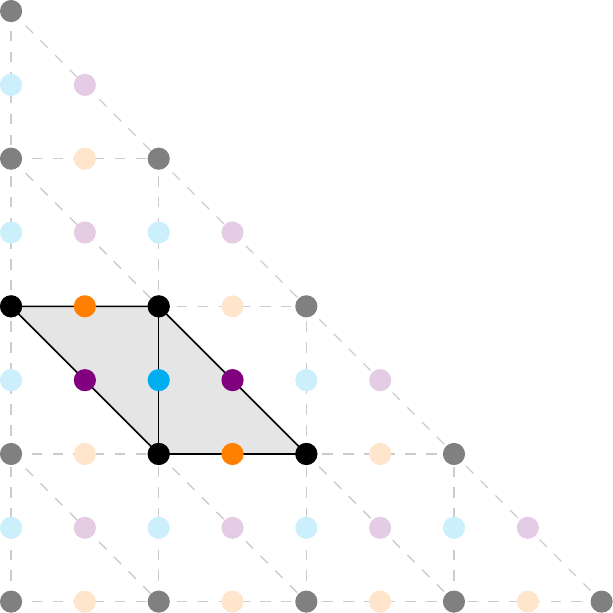}
}}
\caption{Stencils for vertex and edge DoFs of P2 finite elements, visualised as highlighted compact support of the DoFs.}
\label{fig:stencils_P2dofs}
\end{figure}

When solving very large systems of equations obtained from the discretisation of PDEs,
the memory consumption of matrix-based implementations may be impracticably high, 
even when using appropriate sparse matrix formats.
\TERRAN therefore employs matrix-free methods based on stencils (\cf \Fig{fig:stencils_P2dofs}).
Stencil operations additionally allow for efficient parallel kernels to carry out matrix-vector multiplications or point-wise smoothers.

The stencil approach is illustrated in \Fig{fig:stencils_P2dofs} for second-order triangular finite elements.
In this approach, the regularity of the grid is exploited by storing the non-zero elements of a matrix row as stencil entries for the corresponding unknown and its neighbours.
\Fig{fig:vertexDof} shows the stencil associated with a micro-vertex DoF by highlighting the DoFs within the compact support of this vertex.
Figs.~\ref{fig:diagEdgeDoF}--\ref{fig:vertEdgeDoF} display the stencils associated with diagonal, horizontal, and vertical micro-edge DoFs, respectively.

In \TERRAN a matrix is never fully stored in memory, except possibly on coarse levels of a multigrid hierarchy. 
On finer refinement levels, only the result of a matrix-vector multiplication is computable and may be stored in another vector. 
The implementation of the matrix-vector multiplication is up to the developer.
In conventional finite-element solvers this is usually done through quadrature loops which integrate the bilinear forms of the weak form on-the-fly or by using pre-computed local element matrices. 
In the special case of linear PDEs with constant coefficients, only one constant stencil has to be stored for each primitive.
An example for the matrix-vector multiplication implementation in this case is given in \Alg{algo:apply}. 

\begin{algorithm}[h]
    \caption{Sparse operator application $x_{dst} := A x_{src}$ with overlapping communication (\ie, all sends are non-blocking)}
    \label{algo:apply}
    \begin{algorithmic}[1]
        \Procedure{apply-stencil}{$stencil, x_{src}, x_{dst}$}
            \State $x_{src}$: send macro-edge unknowns $\rightarrow$ macro-vertex halos (a)
            \State $x_{src}$: send macro-face unknowns $\rightarrow$ macro-edge halos (b)
            \State $x_{src}$: send macro-vertex unknowns $\rightarrow$ macro-edge halos (c)
            \State $x_{src}$: send macro-edge unknowns $\rightarrow$ macro-face halos (d)
            \State $x_{src}$: wait and recv (a)
            \ParFor{macro-vertex}
            \State on macro-vertex: $x_{dst} \gets$ apply-stencil-macro-vertex($x_{src}$)
            \EndParFor
            \State $x_{src}$: wait and recv (b) and (c)
            \ParFor{macro-edge}
            \State on macro-edge: $x_{dst} \gets$ apply-stencil-macro-edge($x_{src}$)
            \EndParFor
            \State $x_{src}$: wait and recv (d)
            \ParFor{macro-face}
            \State on macro-face: $x_{dst} \gets$ apply-stencil-macro-face($x_{src}$)
            \EndParFor
        \EndProcedure
    \end{algorithmic}
\end{algorithm}

Note again that the computations in the interior of a primitive are independent of other primitives. 
A communication step is required only before primitives of the next dimension are processed.
Thus the work in each primitive may be distributed across processes.

\subsection{Iterative Solvers}
The basic linear algebra routines as discussed in \Sect{subsec:discretizations} and \Sect{subsec:matrixfree} allow the implementation of iterative solvers for linear systems. 
On coarse grids we may employ Krylov or direct solvers.
These solvers can be implemented directly or via 
an interface to external libraries like PETSc~\cite{Petsc1997}.
The preconditioned conjugate gradient 
method implemented in \TERRAN can \eg be 
used for symmetric and positive definite problems.
For indefinite symmetric problems, the preconditioned
minimal residual 
method is typically applied.

For large problems, geometric multigrid solvers are employed. 
Multigrid methods are essential, since they exhibit asymptotically optimal complexity and
thus become superior when progressing to extreme scale~\cite{Gmeiner2016QuantitativeStokes}.
Interpolation and restriction operators between levels are required 
for multigrid and may be realised by matrix-vector multiplications with non-square matrices. 
Note that the most efficient smoothers for multigrid often require pointwise updates and thus
access to the diagonal entries of the stiffness matrix.
This access is provided by the stencil paradigm, but it might not be easy to accomplish in alternative matrix-free methods when only the matrix-vector-multiplication is realised.
In \TERRAN, multigrid can be used as a solver or as a preconditioner in a Krylov solver.

Two major categories of refinement in multigrid methods are h- and p-refinement~\cite{elman2014finite}.
h-refinement describes a refinement process in that the same finite element discretisation is employed on two meshes with different geometric refinement levels.
The combination of different orders of finite element discretisations, \eg P1 and P2 elements, allows to employ p-refinement. 
Here the interpolation and restriction is performed
between different discretisations on the same grid, \ie, on the same geometric refinement level. 
Combining both approaches allows the implementation of geometric multigrid solvers with hp-refinement.
For the solution of large saddle-point problems, an all-at-once multigrid solver using pointwise 
inexact Uzawa smoothers is available~\cite{John2016}.

\section{Example Applications\label{sec:Results}}
The following applications are chosen as a proof of concept for the
software architecture and design aspects.

\subsection{Stokes Flow}
\begin{figure}[ht]
  \centering
   \subfigure[velocity magnitude and stream lines \label{fig:stokesflowporousvelocity}]{
   \resizebox*{0.9\textwidth}{!}{
      \includegraphics[width=\textwidth]{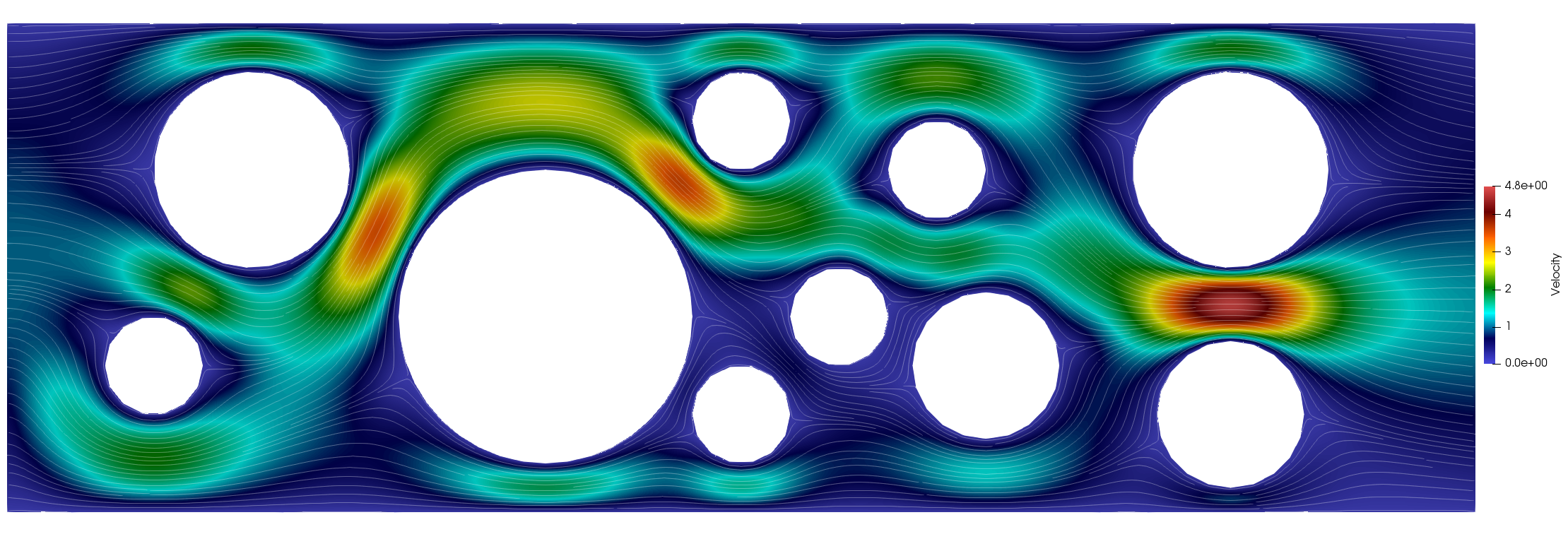}
   }}%

   \vspace{\baselineskip}
   \subfigure[pressure field and macro-faces \label{fig:stokesflowporouspressure}]{
   \resizebox*{0.9\textwidth}{!}{
      \includegraphics[width=\textwidth]{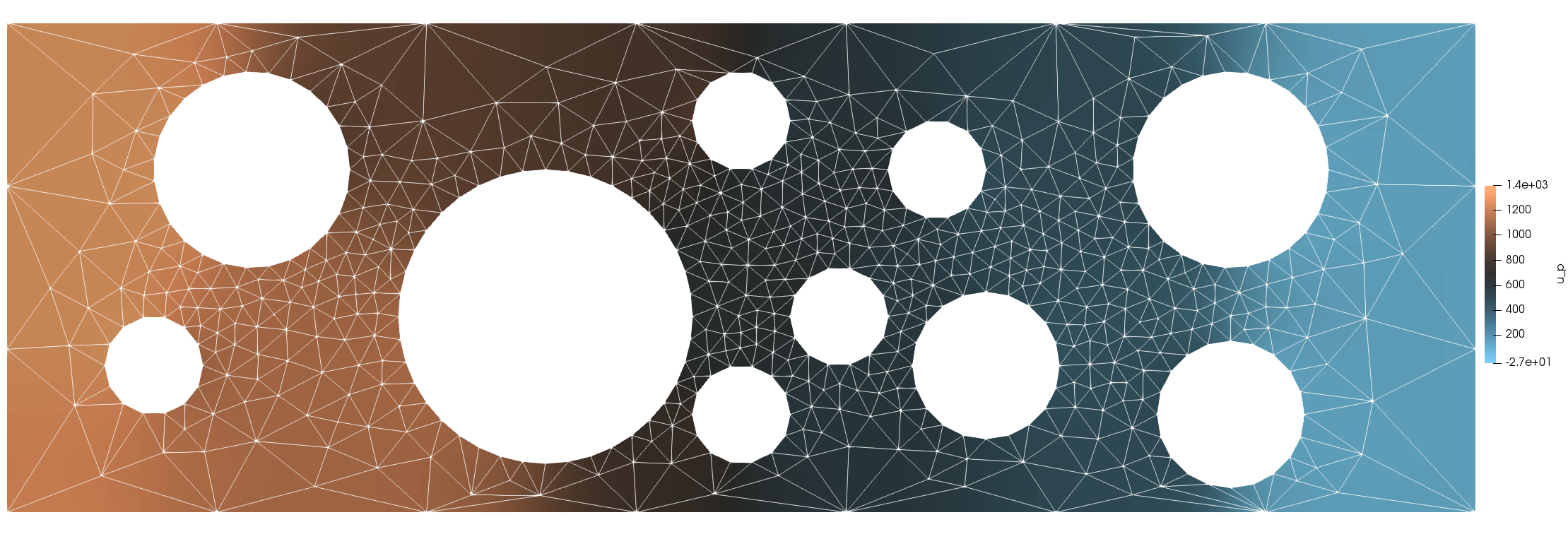}
   }}%
  \caption{Stokes flow through a porous structure with a parabolic inflow profile at the left and outflow at the right boundary.\label{fig:stokesflowporous}}
\end{figure}

In the first example an isoviscous and incompressible Stokes flow through a porous structure is considered (\cf~\Fig{fig:stokesflowporous}) as modelled by 
\begin{align*}
    -\Delta\mathbf{u} + \nabla{p} &= 0,\\
    \mathop{\rm\,div}\mathbf{u} &= 0.
\end{align*}
A parabolic inflow profile is prescribed at the left, no-slip conditions at
the top and bottom, and Neumann outflow boundary conditions at the right
boundary of the domain.
As a stable discretisation, the P2-P1 pairing (\emph{Taylor-Hood}~\cite{elman2014finite}) is used, \ie, the velocity is discretised by
second- and the pressure by first-order finite elements.  The linear system resulting from a discretisation with 
refinement level~4 is directly solved by an $LU$ decomposition using the PETSc interface of \TERRAN.
\Fig{fig:stokesflowporous} shows the velocity magnitude and pressure field as well as the coarse grid
structure of the underlying mesh.

\subsection{Energy Transport}
\begin{figure}[ht]
    \centering
   \subfigure[\label{fig:geophysicsplume00}]{
   \resizebox*{0.46\textwidth}{!}{
      \includegraphics[width=\textwidth]{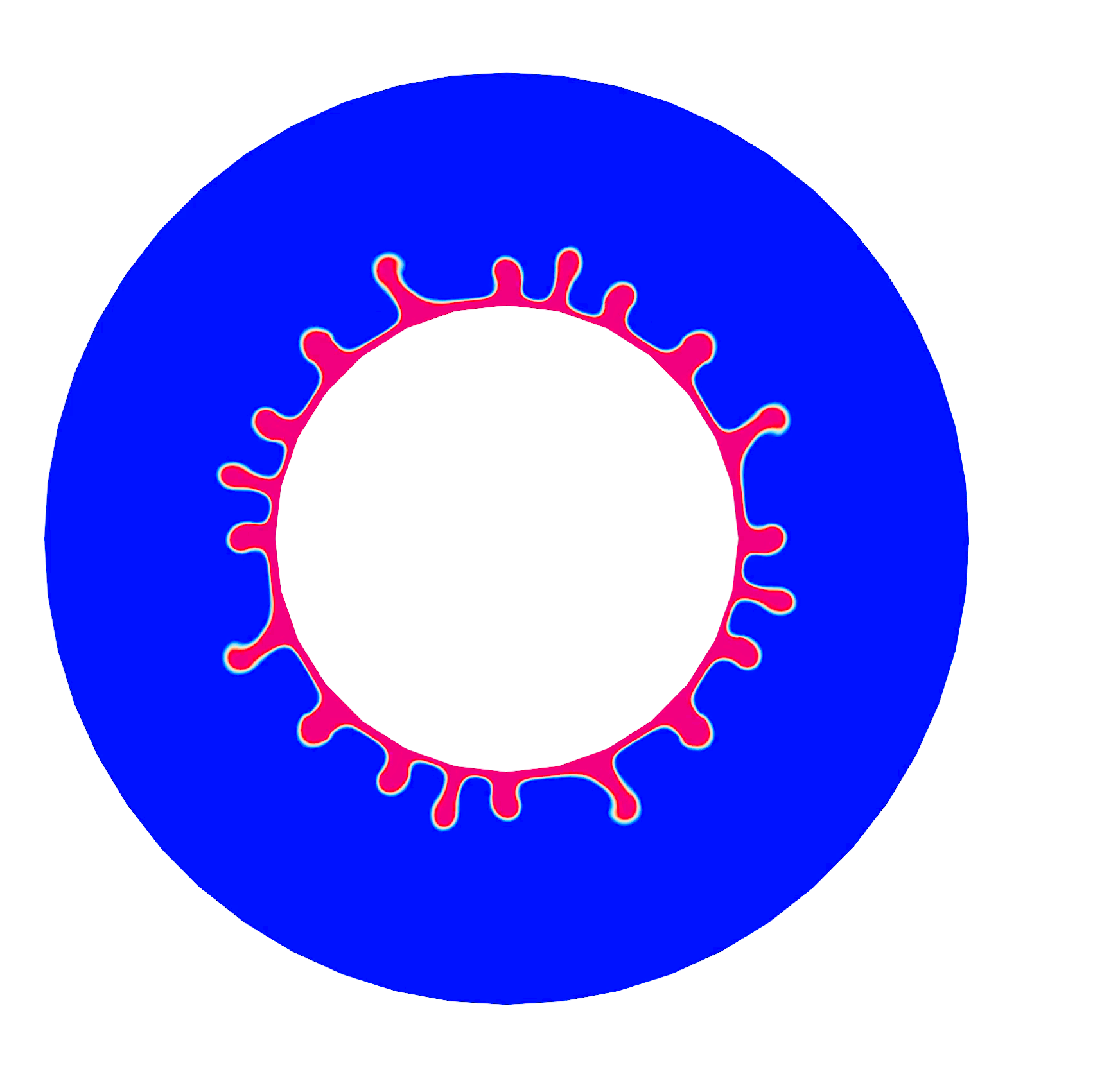}
   }}%
   \subfigure[ \label{fig:geophysicsplume01}]{
   \resizebox*{0.46\textwidth}{!}{
      \includegraphics[width=\textwidth]{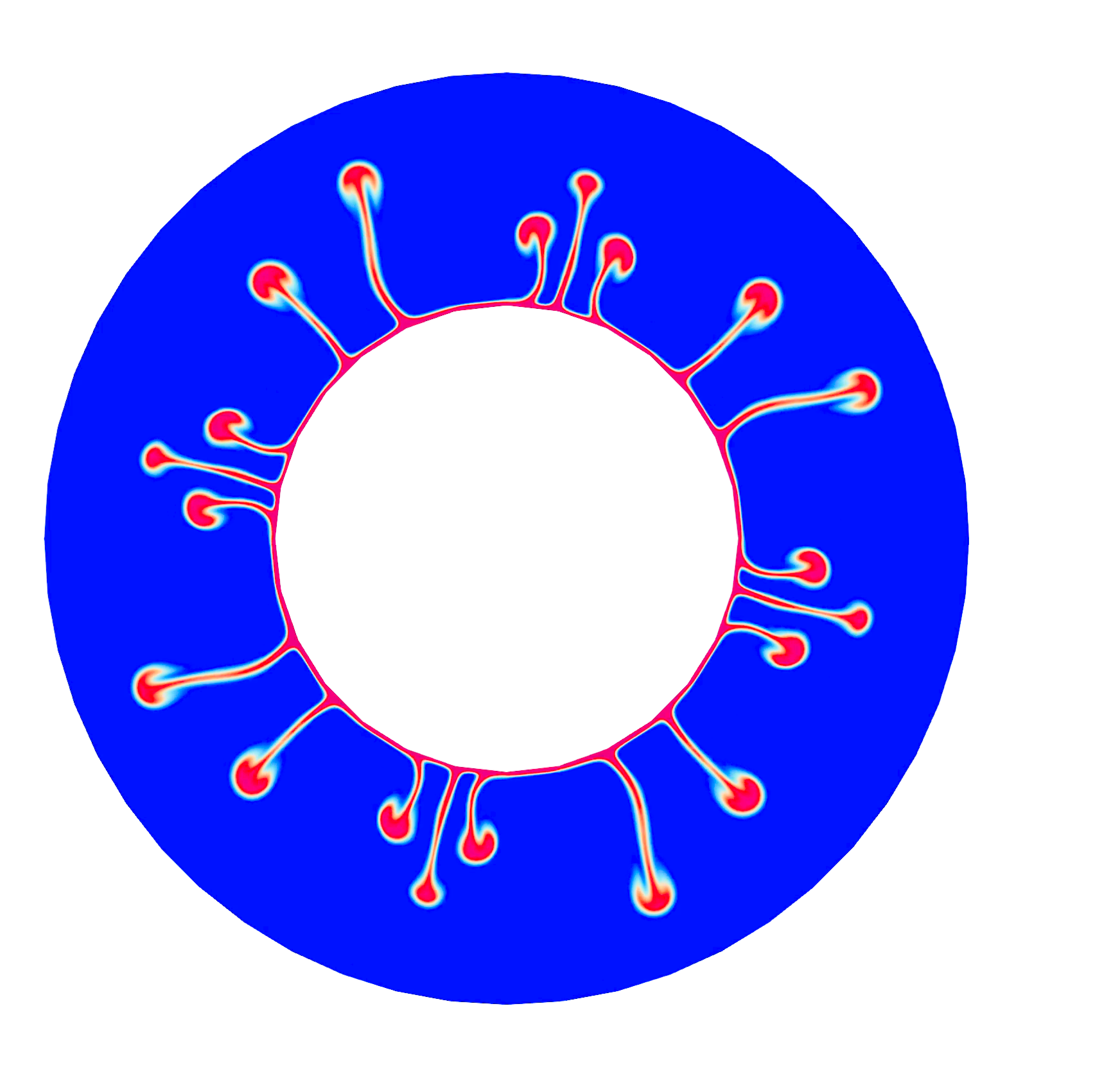}
   }}%

   \subfigure[\label{fig:geophysicsplume02}]{
   \resizebox*{0.46\textwidth}{!}{
      \includegraphics[width=\textwidth]{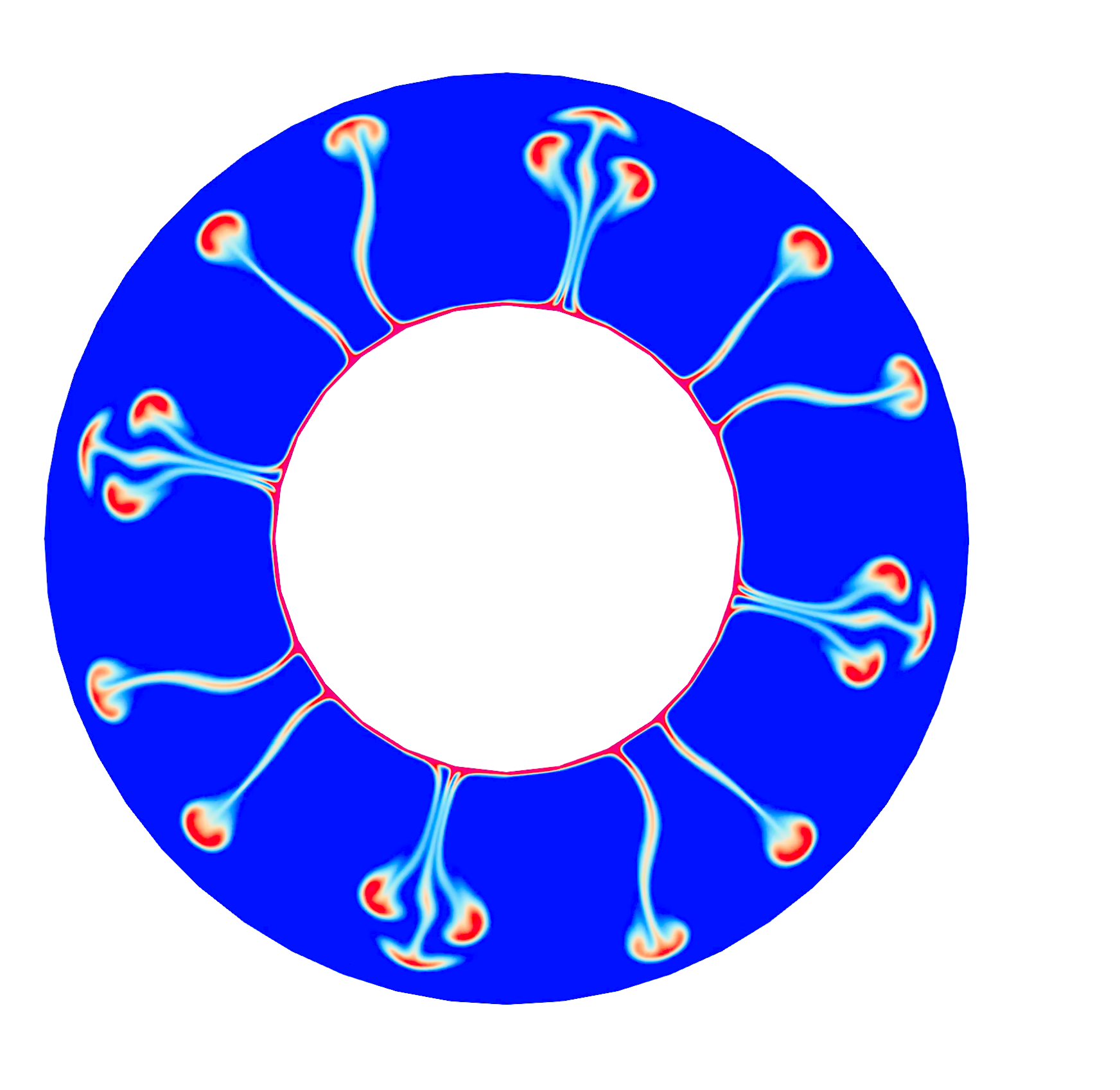}
   }}%
   \subfigure[ \label{fig:geophysicsplume03}]{
   \resizebox*{0.46\textwidth}{!}{
      \includegraphics[width=\textwidth]{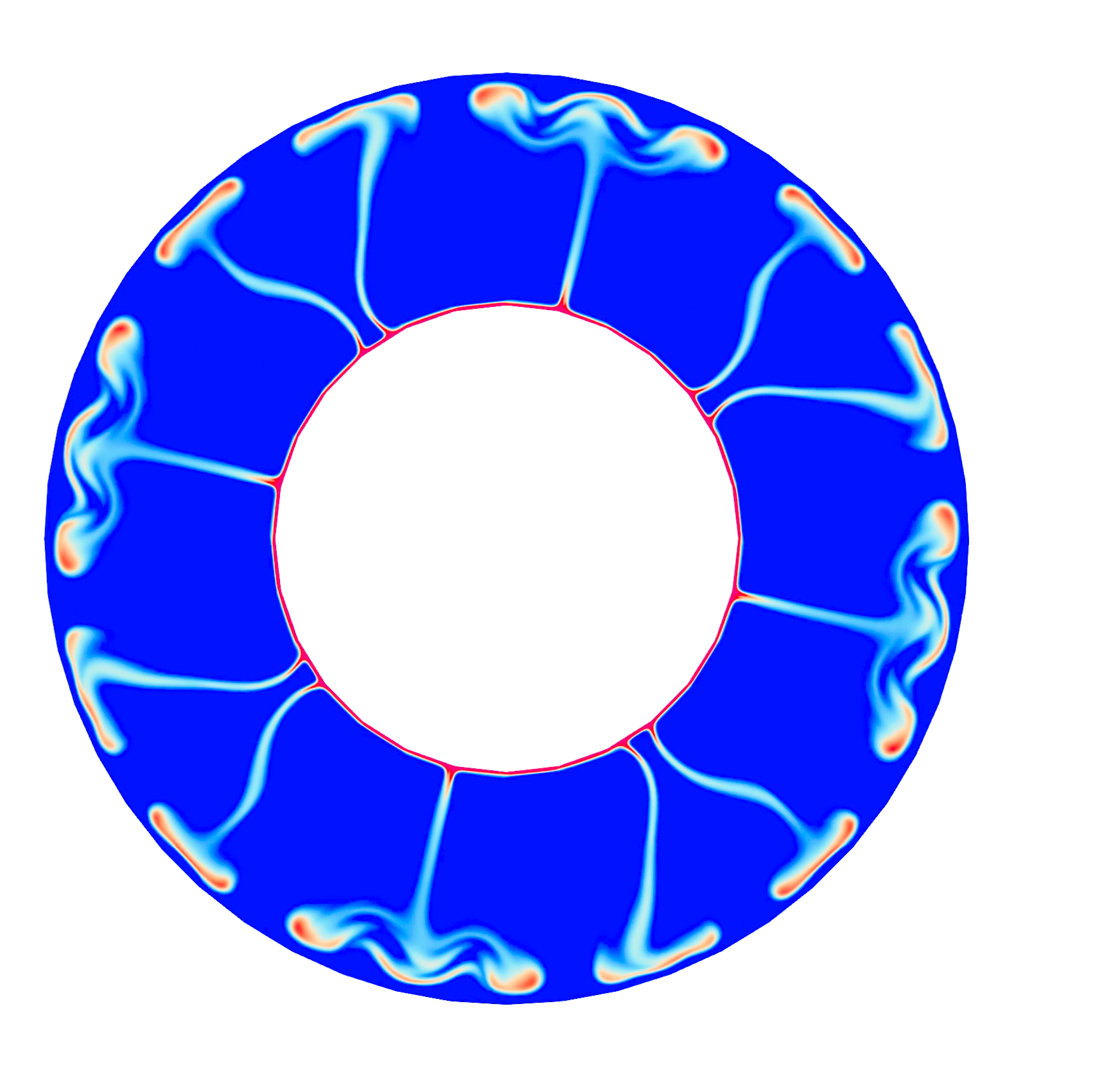}
   }}%
    \caption{Temperature transport modelled by Stokes flow coupled with a convection-diffusion equation on an annulus
      shaped domain with a total number of $500\,000$ time steps.\label{fig:geophysicsplume}}
\end{figure}

In the second example, the temperature driven convection
from the inner to the outer boundary in an annulus is
simulated as a simplified model of Earth mantle convection (\cf~\Fig{fig:geophysicsplume}).  

The process is described by the isoviscous,
incompressible Stokes equations for velocity and pressure,
coupled with a convection-diffusion equation for the
temperature. 
The governing coupled system of equations
for the velocity $\mathbf{u}$, pressure $p$, and temperature $T$ is given by
\begin{align*}
    -\Delta\mathbf{u} + \nabla{p} &= -\mathrm{Ra}\,T\,\widehat{\mathbf{r}}\\
    \mathop{\rm\,div}\mathbf{u} &= 0 \\
    \partial_t T + \mathbf{u}\cdot\nabla{T} &= \mathrm{Pe}^{-1}\Delta{T}.
\end{align*}
Here $\mathrm{Ra}$ represents the dimensionless Rayleigh number, 
$\mathrm{Pe}$ the P{\'{e}}clet number, 
and $\widehat{\mathbf{r}}$ an outward pointing radial vector.
The inverse of the P{\'{e}}clet number is very small, \ie, $\mathrm{Pe}^{-1} \approx 0$,
such that the model is dominated by numerical diffusion. This diffusion becomes smaller with decreasing mesh size.
At the boundaries, homogeneous Dirichlet conditions are employed for the velocity~$\mathbf{u}$.
For the dimensionless temperature a value of $T=1$ is prescribed at the inner boundary and of $T=0$ at the outer boundary.

Here the Stokes system is discretised by equal order P1-P1
finite elements together with a Petrov-Galerkin pressure (PSPG)
stabilisation~\cite{Brezzi1988} for the pressure.
The convection is realised by a finite volume discretisation.
To advance the system in time, the Stokes system and
the convection-diffusion equation are treated with different time steps.
The Stokes system is solved 
only after every three time steps of updating the convection-diffusion equation.
The pressure and velocity are simultaneously solved in a monolithic 
multigrid algorithm, similar to~\cite{John2016}.
This has been proposed in the classical multigrid
literature~\cite{brandt2011multigrid} as the most efficient approach to 
solving the Stokes system.
The temperature is advanced
via an explicit upwind scheme using the previously computed velocity field.

The partitioned annulus domain contains $256$ macro-faces and $2\,145$ unknowns per macro-face on the finest refinement level~6, \ie, in total more than half a million DoFs.

\section{Conclusion and Outlook\label{sec:Conclusion}}  
In this article, the design principles of the generic new 
finite-element framework \TERRAN are presented.
By design, \TERRAN supports higher-order finite elements on a hierarchically discretised domain.
It provides scalable, parallel data structures, sophisticated load balancing,
an abstract memory layout, and a layered communication concept.
The basic design will also support adaptivity and asynchronous execution
as needed  when progressing to extreme-scale computing.
The approach to achieve high single-node performance and good scalability within \TERRAN is based on the combination of an unstructured topology with structured data inside the macro-primitives. 
The data structures can be used to implement efficient matrix-free methods and the basic numerical building blocks for iterative solvers. 
The stencil approach allows for the scalable and memory-aware implementation of point-wise smoothers that are essential to employ fast geometric multigrid solvers.

The concept is designed towards a seamless transition to three-dimensional simulations building on tetrahedral macro-primitives and is currently extended.
In future work the scalability and performance of \TERRAN will be examined in more detail.
A major focus lies on the implementation of massively parallel geometric multigrid
solvers for the Stokes system and their application to 
large-scale geophysical simulations.
Beyond these first examples, the framework is also well suited for other
applications that require large-scale simulations with
finite elements, such as in electromagnetics or fluid dynamics.
\section*{Acknowledgements}
This work was partly supported by the German Research Foundation through
the Priority Programme 1648 "Software for Exascale Computing" (SPPEXA) and by grant WO671/11-1.

\bibliography{main_journal_JPED2018.bbl}


\label{lastpage}

\end{document}